\documentclass[review,12pt,3p,authoryear]{elsarticle} 

\usepackage{amsmath,amssymb}

\usepackage{xcolor}
\usepackage[figurename=Fig.]{caption}
\def\citeapos#1{\citeauthor{#1}'s (\citeyear{#1})}
\journal{International Journal of Multiphase Flow}
\usepackage{multirow} 
\usepackage{bm}
\usepackage{hyperref}
\hypersetup{
urlcolor=blue,
}
\usepackage{fancyhdr}
\begin{document}

\begin{frontmatter}

\title{Motion of finite-size spheres released in a turbulent boundary layer}

\author[mymainaddress]{Yi Hui Tee\corref{mycorrespondingauthor}}
\cortext[mycorrespondingauthor]{Corresponding author}
\ead{teexx010@umn.edu}
\author[mymainaddress,mysecondaryaddress]{Diogo Barros}
\author[mymainaddress]{Ellen K. Longmire}
\address[mymainaddress]{Aerospace Engineering and Mechanics, University of Minnesota, Minneapolis, USA} 
\address[mysecondaryaddress]{Aix Marseille {Universit\'e}, CNRS, IUSTI,  
 Marseille, France}

\begin{abstract}
Individual magnetic wax spheres with specific gravities of 1.006, 1.054 and 1.152 were released from rest on a smooth wall in  water at friction Reynolds numbers, $Re_{{\tau}} = 680$ and 1320 ({sphere diameters} $d^+ = 58$ and 122 viscous units, respectively).
Three-dimensional tracking was conducted to understand the effects of turbulence and wall friction on sphere motions. 
Spheres subjected to sufficient mean shear initially lifted off of the wall before descending back towards it.
These lifting spheres translated with the fluid above the wall, undergoing saltation or resuspension, with minimal rotations about any axis.
By contrast, spheres that did not lift off upon release mainly slid along the wall. 
These denser spheres lagged the fluid more significantly due to greater wall friction.
As they slid downstream, they began to roll forward after which small repeated lift-off events occurred.
These spheres also rotated about both the streamwise and wall-normal axes. 
In all cases, the sphere trajectories were limited to the buffer and logarithmic regions, and all wall collisions were completely inelastic.
In the plane parallel to the wall, the spheres migrated in the spanwise direction about $12\%$ of the streamwise distance traveled suggesting that spanwise forces are important.
Variations in sphere kinematics in individual runs were likely induced by high and low momentum zones in the boundary layer, vortex shedding in the sphere wakes, and wall friction.
The repeated lift-offs of the forward rolling denser spheres were attributed to a Magnus lift. 

\noindent {\textcopyright  2020. This manuscript version is made available under the CC-BY-NC-ND 4.0 license; \url{http://creativecommons.org/licenses/by-nc-nd/4.0/}.}
\end{abstract}


\end{frontmatter}

\section{Introduction}

Particle-laden turbulent flows occur in many applications ranging from industrial processes to the environment such as the pollutant particles in atmosphere, rivers and oceans \citep[e.g.,][]{olsen1982pollutant, law2014microplastics, castaneda2014microplastic}.
In a wall-bounded flow, the particle motion is complicated by interactions with both multiple scales in the turbulent flow and the wall itself.
The presence of coherent structures such as the alternating high and low momentum regions associated with ejection and sweep events in the near-wall regions \citep[]{wallace1972wall} can induce complex particle-turbulence interactions.
When a particle is larger than the smallest fluid eddies, it can experience variations in shear and normal forces around its circumference.
Additionally, wall friction and restitution can affect both the translation and rotation of the particles. 
Depending on the surrounding conditions, particles can either collide with or lift off from the wall or slide or roll along it.
All these effects can significantly impact particle resuspension, deposition and transport which are all very crucial in modeling the particle response in real applications.

Various early experiments examined particle dynamics in turbulent open-channel flows.
Among others, \cite{sutherland1967proposed} investigated how grains in a sediment bed were brought into motion by the fluid.
He proposed an entrainment hypothesis whereby strong turbulent eddies could disrupt the viscous sublayer and lift the grain off of the bed.
\cite{francis1973experiments} and \cite{abbott1977saltation} observed rolling, saltation and suspension behavior of heavy grains transported over a planar rough bed.
Meanwhile, \cite{sumer1978particle} and \cite{sumer1981particle} observed that a sand-coated wax sphere with diameter of approximately 30 viscous units ($d^+$) propagated upwards and downwards repetitively throughout its trajectory in both rough and smooth beds. 
Hereafter, the symbol $^+$ is used to denote quantities normalized by the friction velocity ($u_\tau$) and the kinematic viscosity ($\nu$) of water at $20^\circ \text{C}$. 

{To better understand particle-turbulence interactions in wall-bounded flows, direct visualization techniques were incorporated in investigations by \cite{rashidi1990particle}, \cite{kaftori1995particle, kaftori1995particleb}, \cite{ninto1996experiments}, \cite{van2013spatially} and \cite{ebrahimian2019dynamics} to name a few}.
These studies concluded that particle resuspension and deposition events in the near-wall regions were strongly influenced by coherent flow structures. 
Specifically, \cite{van2013spatially} reported that in his time-resolved particle image velocimetry (PIV) and particle tracking velocimetry (PTV) experiments, all lift-off events of polystyrene beads ($d^+=10$) at friction Reynolds number, $Re_{{\tau}}=435$ were due to ejection events generated by passing vortex cores and positive shear. 
Once lifted beyond the viscous sublayer, the particles either stayed suspended in the fluid or saltated along the wall depending on the type of coherent structures that they encountered.
\cite{ebrahimian2019dynamics}, who studied glass beads ($d^+=6.8$) in channel flow with $Re_{{\tau}}=410$, found the strongest bead accelerations were correlated with fluid ejections and occurred at a wall-normal location of $y^+=30$.
Sweep motions contributed to streamwise decelerations of beads closer to the wall ($y^+<20$).
In addition, \cite{rashidi1990particle} and \cite{kaftori1995particle, kaftori1995particleb}, who investigated spanwise motions in dilute particle suspensions in turbulent boundary layers, concluded that the small particles tended to accumulate in the low-speed zones where ejection events were prominent.  

A key factor in particle saltation and resuspension is the lift force acting on a particle. 
Sphere lift can result from mean shear \citep[]{saffman1965lift}, individual vortices \citep[]{auton1987lift}, or turbulence \cite[]{sutherland1967proposed} in the fluid, from solid body rotation \citep[]{magnus1853ueber} or a combination of both (see reviews by \citeauthor*{loth2008lift}, \citeyear{loth2008lift} and \citeauthor*{shi2019lift}, \citeyear{shi2019lift}).
When a sphere is close to a wall, the surrounding flow field is distinctly different from that in an unbounded flow.
\cite{hall_1988} and \cite{mollinger1996measurement}, who measured the fluid-induced forces acting on a particle fixed to the wall in a turbulent boundary layer using force transducers, reported strong positive mean force in the wall-normal direction. 
\citeapos{hall_1988} experimental data showed that, for $3.6 < d^+ < 140$ and particle Reynolds number, $6.5<Re_p<1250$, the normalized mean force could be approximated by $F_{y}^+=(20.90\pm1.57)(d^+/2 )^{2.31\pm0.02}$.
On the other hand, \citeapos{zeng2008interactions} fully-resolved direct numerical  simulation (DNS) of the flow over a fixed sphere located a distance above the wall showed the opposite.
The mean wall-normal forces, evaluated by integrating the non-dimensional pressure and viscous stress around the sphere surfaces, were negative for all spheres with $3.56\leq d^+\leq 24.94$ centered at a wall-normal location of {$y^+=17.81$} at $Re_{{\tau}}=178.12$.
Tomographic PIV performed by \cite{van2018experimental} also suggested a negative wall-normal contribution on a tethered sphere with $d^+=50$ centered at $y^+=43$ above the wall at $Re_{{\tau}}=352$ due to the sphere wake tilting away from the wall.
These results imply that the net wall-normal force including the lift force can vary significantly depending on the gap between sphere and wall.

While the presence of a wall can affect the lift force acting on a non-rotating particle considerably, it can also cause the sphere to roll due to frictional torque.
In addition, the hydrodynamic torque generated by strong velocity gradients, vorticity and turbulent fluctuations can also cause the sphere to rotate \citep[e.g.,][]{saffman1965lift, cherukat1999computational, bagchi2002effect, bluemink2008sphere}.
In turbulent boundary layer flows, \cite{white1977magnus} and \cite{nino1994gravel2}, among others, evaluated the significance of Magnus lift on saltating particles based on comparisons between theoretical and experimental particle trajectories.
They concluded that Magnus lift could be a non-negligible part of the overall particle lift force.
However, rotation was not quantified in either study.

In many previous numerical simulations of wall-bounded flows, particles were modeled as point-masses with no volume and thus no rotation \citep[e.g.,][]{pedinotti1992direct, dorgan2004simulation, soldati2009physics}.
Particle-resolved simulations are relatively limited due to the high computational cost and challenges in getting fine resolution \citep[]{balachandar2010turbulent}. 
In most of the particle-laden wall-bounded flow simulations, even though finite-size effects were taken into account, effects due to particle rotation were neither considered nor discussed explicitly \citep[e.g.,][]{pan1997numerical, zeng2008interactions, fornari2016effect}.
Several studies including those from \cite{zhao2011particle}, \cite{ardekani2019turbulence} and \cite{peng2019direct}, reported that particle rotation can induce significant effects on the turbulence modulation and should not be neglected. 

Several experimental techniques have been proposed to measure sphere translation and rotation simultaneously.
For example, \cite{zimmermann2011tracking} and \cite{mathai2016translational} extracted the sphere position and absolute orientation by comparing the sphere images with unique patterns captured from two perpendicular cameras to a database of synthetic projections.
Meanwhile, \cite{klein2013simultaneous} and \cite{barros2018measurement} tracked the markers embedded within or painted over the surface of a sphere using multiple cameras.
Then, the angular velocity was computed based on the optimal rotation matrix that best aligned the tracked markers \citep[]{kabsch1976solution,kabsch1978discussion}.

As presented above, in a particle-laden wall-bounded turbulent flow, the particle-turbule-nce and particle-wall interactions are complicated to resolve.
In the context of particle motion, most attention has been devoted to the two-dimensional translational behavior of particles over short streamwise distances, with little attention to rotation.
In order to comprehend the particle dynamics more fully, we track individual spheres over significant streamwise distances while resolving all components of translation and rotation. 
Multiple sphere densities and flow Reynolds numbers are considered to study the effects of specific gravity and mean shear on sphere motions.
\citeapos{barros2018measurement} methodology is adapted to the requirements of the current experimental setup to reconstruct both the sphere position and orientation. 
The paper is organized as follows: Section \ref{Methodology} describes the experimental setup, parameters, reconstruction of particle motion and uncertainty analysis; in Section \ref{Results}, we present the results on sphere translational and rotational kinematics, and discuss how they are affected by both turbulence and the bounding surface; the concluding remarks are summarized in Section \ref{Section4}.

\section{Methodology}
\label{Methodology}

\subsection{Experimental Setup}
\label{Setup}
The experiments were conducted in a recirculating water channel facility at the University of Minnesota.
The channel test section, which is constructed of glass, is 8 m long and 1.12 m wide. 
A 3 mm cylindrical trip-wire was located at the entrance of the test section to trigger the development of a turbulent boundary layer along the bottom wall. 
Further details on the facility can be found in \cite{gao2011evolution}.
Hereafter $x$, $y$ and $z$ define the streamwise, wall-normal, and spanwise directions, respectively. 

\begin{figure*}[]
        \centering

    	 \includegraphics[trim={0 .4cm 0 0 cm },clip,width=0.96\textwidth]{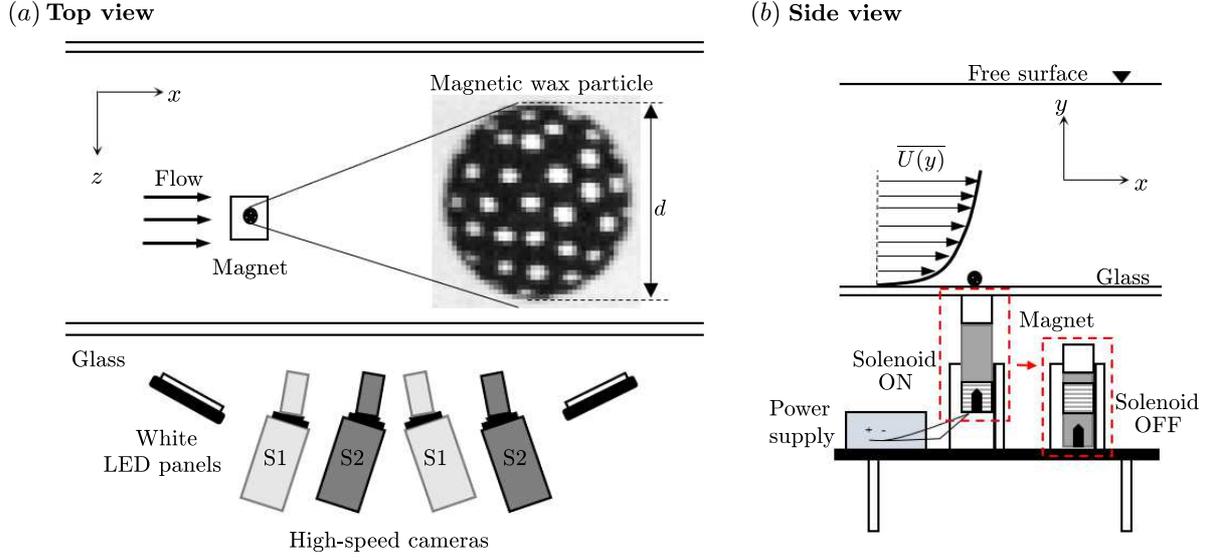} %
	 \caption{Experimental setup. ($a$) Top view: two pairs of high-speed cameras (S1 and S2) aligned in stereoscopic configurations for capturing the trajectory and rotation of a marked sphere over a long field of view. Inset: example of sphere captured in grayscale with diameter, $d$ spanning 43 pixels or 6.35 mm. ($b$) Side view: sphere on smooth glass wall in a turbulent boundary layer. Red boxes illustrate the mechanism used in holding (solenoid ON) and releasing (solenoid OFF) the sphere, respectively.} 

		\label{fig:1}

\end{figure*}

To achieve a repeatable and controllable initial condition, magnetic spheres, with diameter ($d$) of $6.35\pm0.05$ mm, molded from a mixture of blue machinable wax (913.7 $\text{kg m}^\text{-3}$) and synthetic black iron oxide (5170 $\text{kg m}^\text{-3}$) were used. 
Here, a small piece of wax was melted inside two hemisphere molds and different amounts of iron oxide particles were added to control the sphere density.
Then, the two molds were brought together and chilled immediately to solidify the liquid wax. 
The resulting spheres were black and opaque.
Small markers were painted at arbitrary locations all over the sphere surface using a white oil-based pen for good image contrast (see Fig.\ \ref{fig:1}$a$).
Both the mean inter-marker spacing and mean marker diameter were approximately 0.6 mm.

For each run, a given sphere was held statically on the smooth glass wall in the boundary layer by a magnet at a location 4.2 m downstream of the trip wire and 0.3 m (approximately 4 boundary layer thicknesses, $\delta$) away from the nearest sidewall based on the sphere centroid.
This location will be considered as the origin in $x$ and $z$, with the bottom wall as $y=0$. 
A DC 12V 2A push-pull type solenoid from Uxcell was used to hold the cubic neodymium magnet (N40) and the sphere in position. 
When activated, the solenoid held the plunger such that the magnet which was connected through a slider will be flush with the outer channel wall. 
By switching off the power supply, both the plunger and the magnet were retracted with the help of gravity (see Fig.\ \ref{fig:1}$b$). 
As the magnet moved away from the bottom wall, the sphere was released, allowing it to propagate with the incoming flow. 
A coarse screen was located at the end of the test section to capture the sphere and prevent it from recirculating around the channel. 

Two pairs of Phantom v210 high-speed cameras from Vision Research Inc.\ were arranged in stereoscopic configurations to track the sphere in three-dimensional (3D) space over a relatively long field of view. 
The angle between the two stereoscopic cameras was set to approximately $30^\circ$ for both camera pairs.
The cameras were positioned with a streamwise overlapping distance of approximately two particle diameters in between their fields of view. 
All cameras were fitted with 105 mm Nikon Micro-Nikkor lenses with aperture $f/16$. 
Scheimpflug mounts were added to all cameras so that the images were uniformly focused across the fields of view. 
Three white LED panels positioned above the cameras illuminated the domain considered.
Prior to running the experiments, the optical system was calibrated by displacing a two-level plate (LaVision Type 22) across nine planes in the spanwise direction for both camera pairs. 
A third order polynomial fit was obtained for each plane from both cameras pairs using the calibration routine of Davis 8.4 (LaVision GmbH) to generate the mapping function of the volumetric calibration respectively. 
The root mean square (r.m.s.) error of the grid point positions was between 0.05 and 0.1 pixels indicating an optimal fit.
Image sequences were captured at a sampling frequency of 480 Hz with image resolution of 1280 by 800 pixels, \textcolor{black}{pixel depth of 12-bit, and camera pixel size of 20 $\mu$m}.

\subsection{Experimental Parameters}
\label{Parameters}
To understand the effect of turbulence, the experiments were conducted at two flow Reynolds numbers.
Here, the mean flow statistics of the unperturbed turbulent boundary layers at the initial sphere location were determined from planar PIV measurements in streamwise wall-normal planes.
A New Wave Solo II Nd:YAG 532 nm double-pulsed laser system with pulse energy of 30 mJ was used for illumination.
The laser sheet illuminated through the bottom glass wall had a thickness of 1 mm. 
The flow was seeded with silver-coated hollow glass spheres from Potters Industries LLC with an average diameter and density of 13 $\mu$m and 1600 $\text{kg m}^\text{-3}$, respectively. 
At both flow Reynolds numbers, 2000 image pairs were acquired using a TSI Powerview Plus 4MP 16-bit PIV camera \textcolor{black}{with camera pixel size of 7.4 $\mu$m.
The image pairs were} captured at a sampling frequency of 1.81 Hz with image resolution of 2048 by 2048 pixels.

\begin{figure}[h]
		
        \centering

        {\includegraphics[width=1\textwidth]{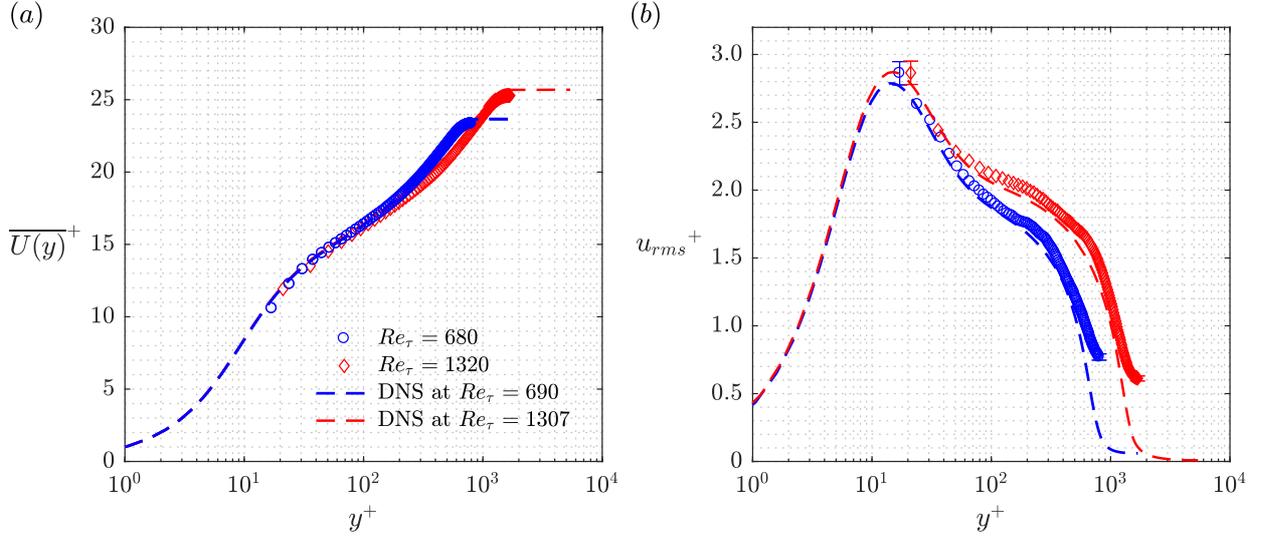}}

\caption{Statistics of the unperturbed turbulent boundary layers. (${a}$) Mean streamwise fluid velocity and ($b$) r.m.s.\ of streamwise velocity  fluctuations. Lines:  DNS profiles from \cite{jimenez2010turbulent}. Symbols: PIV data. Blue: $Re_\tau=680$; Red: $Re_\tau=1320$. Error bars in ($b$) indicate the measurement uncertainties. 
	}
		\label{fig:2}
\end{figure}

The PIV images were processed using Davis 7.4 (LaVision GmbH) to obtain the velocity vectors. 
Normalized cross-correlation \citep[]{fincham1997low} with an overlap of 50\% over initial interrogation window sizes of 64 by 64 pixels followed by three passes of 32 by 32 pixels were employed.
All vectors were post-processed with the universal outlier detection criterion \citep[]{westerweel2005universal} to remove spurious vectors.
By applying the Clauser chart method using log-law constant $B=5$ and von K{\'a}rm{\'a}n constant $\kappa=0.41$, $u_\tau$ were estimated to be $0.0092 \text{ and 0.0193 m s}^\text{-1}$ \citep[]{clauser1956turbulent, monty2009comparison}.
At a water depth of 0.394 m under both fluid conditions, the free-stream velocities ($U_\infty$) were 0.215 and 0.488 $\text{m s}^\text{-1}$, with $\delta$ of 73 and 69 mm estimated based on the mean unperturbed streamwise fluid velocity location of $\overline{U(\delta)}=0.99 U_\infty$.
This corresponded to friction and momentum thickness Reynolds numbers of $Re_{{\tau}}=u_\tau \delta/\nu =680$ and 1320 and $Re_{{\theta_m}}=U_\infty \theta_m/\nu =1870$ and 3890, respectively where $\theta_m$ is the momentum thickness.
The spatial resolutions of the computed velocity vectors were 14 and 29 viscous units at $Re_{{\tau}}$ of 680 and 1320, equivalent to 1.51 mm. 
The boundary layer properties are summarized in Table \ref{tab:flow} while the mean flow statistics are plotted in Fig.\ \ref{fig:2}.
The PIV streamwise velocity statistics showed good agreement with the DNS results of \cite{jimenez2010turbulent}.
 
\begin{table}[!]
\caption{Summary of turbulent boundary layer properties.}
\centering
\small
\setlength{\tabcolsep}{15pt}   
\begin{tabular}{c  c  c  c  c }
\hline\hline \\[-4ex]
$U_\infty$ [$\text{m s}^\text{-1}$]  & $u_\tau$ [$\text{m s}^\text{-1}$] & $\delta$ [mm] & $Re_{{\tau}}$ & $Re_{{\theta_m}}$  \\[0.95ex]
\hline \\[-3.7ex]
$0.215$  & $0.0092\pm0.0001$ &  $73\pm2$ & $680\pm20$   & $1870\pm20$   \\
$0.488$   & $0.0193\pm0.0002$ &  $69\pm2$ &  $1320\pm40$ & $3890\pm40$    \\[0.35ex]
\hline\hline
\end{tabular}
\vspace{0.5cm}
\label{tab:flow}
\end{table}

Spheres with $d=6.35\pm0.05$ mm and specific gravities ($\rho_p/\rho_f$) of 1.006 (P1), 1.054 (P2) and 1.152 (P3) were considered, where $\rho_p$ is the sphere density and $\rho_f$ is the fluid density. 
This corresponds to $d^+=d u_\tau / \nu$ of 58 and 122 viscous units, which are comparable to 26 and 46 times the Kolmogorov length scale ($\eta$) in the logarithmic region \citep[]{pope_2000}.
For each sphere, the density was determined based on the settling velocity ($V	_s$) measured from high-speed imaging of the sphere falling in a quiescent fluid where $\rho_p = 3C_D\rho_fV_s^2/4dg + \rho_f$. 
Here, $C_D$ refers to the drag coefficient obtained from the standard drag curve \citep[]{clift2005bubbles} while $g$ is the gravitational acceleration. 
For all spheres, the uncertainty in density was less than 1\%.
Meanwhile, the initial particle Reynolds numbers defined as $Re_p=U_{rel}d/\nu$ were 760 and 1840, where $U_{rel} = 0.122$ and 0.292 $\text{m s}^\text{-1}$ are the relative velocity between particle and mean fluid at the particle center upon release. 
In our cases, Stokes numbers ($Sk_\eta$), expressed as the ratio of the particle's response time, $\tau_p=(\rho_f+2\rho_p)d^2/36\nu\rho_f$  \citep[]{crowe2005multiphase} to the characteristic flow time scale based on Kolmogorov, $\tau_\eta$ \citep[]{pope_2000}, ranged from 59 to 193.
Note that although $|V_s|/U_\infty$ is relatively small for sphere P1, it is very significant for P2 and P3.
Details of the experimental parameters are summarized in Table \ref{tab:1}.
The particles were tracked over a streamwise distance up to $x \approx 5.5\delta$.
For each case considered, $R=10$ trajectories were captured using the same sphere.
\textcolor{black}{For spheres P1 and P3, an additional 55 and 65 runs respectively were completed at both $Re_\tau$ to track the initial sphere lift-off behavior.}

\begin{table}[]
\caption{Summary of experimental parameters.
$\overline{F_y}^*=(\overline{F_y}-F_b)/F_b$ where $\overline{F_y}$ and $F_b$ denote the mean wall-normal fluid-induced force based on \citeapos{hall_1988} expression and the net buoyancy force, respectively.}
\centering
\small
\resizebox{\textwidth}{!}{\begin{tabular}{c  c  c  c  c  c  c c c }
\hline\hline \\[-4ex]
\multirow{2}{*}{$Re_{{\tau}}$}  & $\overline{U(y=\frac{d}{2})}$ &  \multirow{2}{*}{Initial $Re_p$} & \multirow{2}{*}{$d^+$} & \multirow{2}{*}{Sphere} & \multirow{2}{*}{$\rho_p/\rho_f$} & \multirow{2}{*}{$Sk_\eta$} & \multirow{2}{*}{$|V_s|/U_\infty$} & \multirow{2}{*}{Initial $\overline{F_y}^*$} \\
& [$\text{m s}^\text{-1}$]&&&&&& \\[0.95ex]
\hline \\[-3.7ex]
$680$   & $0.122$ & $760$            & $58\pm2$    & P1     & $1.006\pm0.003$    &59   & $0.081$          & $12\pm2$     \\
      & &               &       & P2     & $1.054\pm0.006$       & 60    & $0.43$          & $-0.20\pm0.12$      \\
      &  &              &       & P3     & $1.152\pm0.015$     & 64      & $0.74$          & $-0.74\pm0.04$      \\[0.35ex]
\hline  \\[-3.7ex]
$1320$   & $0.292$ & $1840$           & $122\pm5$   & P1     & $1.006\pm0.003$    & 176   & $0.036$          & $72\pm12$     \\
      &   &              &       & P2     & $1.054\pm0.006$    &  181      & $0.19$          & $3.4\pm0.7$      \\
       &  &               &       & P3     &  $1.152\pm0.015$   &  193       & $0.33$         & $0.45\pm0.24$     \\[0.35ex]
\hline\hline 
\end{tabular}}
\vspace{0.5cm}
\label{tab:1}
\end{table}

\subsection{Reconstruction of Particle Translation and Rotation}
\label{Reconstruction}
Before computing the particle translation and rotation, the grayscale images were first pre-processed using Matlab to isolate the sphere from the background. 
A standard circular Hough Transform routine was applied to locate the sphere.
Next, the background surrounding the sphere was removed by setting the intensity values to 0 (black).
The extracted sphere images were then imported to Davis 8.4.
Here, the images were further processed with 3 x 3 Gaussian smoothing and sharpening filters to increase the dot contrast. 
Pixel intensity values that were less than those of the white dots were set to 0 to isolate the dots from the sphere image. 
In all images, the minimum digital dot size was approximately 2 x 2 pixels.
Subsequently, a 3D-PTV routine based on the volumetric calibration mapping function was implemented to reconstruct the dot coordinates from both camera pairs.

The data sets obtained from the PTV were composed of the 3D coordinates of true and ghost markers and their corresponding 3D velocity vectors. 
Hence, the filtering methodology proposed by \cite{barros2018measurement} was employed to remove the ghost tracks.  
Once the true markers had been determined, the sphere centroid was determined by applying the equation of a sphere. 
Then, a rotation matrix that best aligned the markers of consecutive images was obtained by applying \citeapos{kabsch1976solution} algorithm.
In all runs, at least 8 markers were retained when computing the sphere centroid locations and rotation matrix.
Even though all spheres were captured at 480 Hz, different processing frequencies were used in tracking. 
Depending on the sphere translation speed, the processing frequency was optimized to ensure that marker displacement between images was larger than the disparity uncertainty (see Section \ref{Uncertainty}) while limiting maximum sphere streamwise displacement to less than 10 pixels to avoid false marker pairing.
More details of the reconstruction process can be found in \cite{barros2018measurement}.

\subsection{Uncertainty Analysis}
\label{Uncertainty}
For the PIV measurements, the random errors of the mean velocity vector in both cases, computed based on a 95\% confidence interval, were less than 0.2\% of the local mean velocity.
For the r.m.s.\ of the streamwise velocity fluctuations ($u_{rms}$), the corresponding maximum statistical uncertainty estimated based on chi-square analysis with a 95\% confidence interval was 3\% (see the errorbars plotted in Fig.\ \ref{fig:2}$b$). 
The accuracy of $\delta$ is limited by the PIV vector spatial resolution and this gives an uncertainty of up to 3\% from the estimated value.
Meanwhile, the uncertainty of $u_\tau$ evaluated based on fitting the logarithm law of the wall equation between $y^+>30$ and $y/\delta< 0.2$ was approximately 1\%.
These resulted in an uncertainty of 3\% in the $Re_\tau$ calculation.
Next, the uncertainty of $Re_{\theta_m}$, which was estimated based on the statistical error of the mean velocity data, was around 1\%. 

For the marker reconstruction, the mean disparity error, $\epsilon_{disp}^*$ calculated by projecting the 3D reconstructed markers back to the camera image in Davis 8.4,  was approximately 0.8 px. 
This gives an estimate of the uncertainty in the marker locations due to reconstruction errors \citep{wieneke2008volume}.
To reduce the noise in computing derivatives, the raw position and orientation data were smoothed by a quintic spline \citep{epps2010evaluating}.
The mean uncertainties of the sphere position ($x$, $y$ and $z$) as well as orientation ($\theta_x$, $\theta_y$ and $\theta_z$) were computed based on the r.m.s.\ between the raw and smoothed data \citep{schneiders2017track}.
For sphere position, the uncertainties were 0.51, 0.18 and 0.95 px; for orientation, the values were 1.2, 1.5 and 0.52 px, respectively.
These correspond to mean translational and angular displacement uncertainties of 0.01$d$ and 1.5$^\circ$.
Lastly, the mean uncertainties of the translational sphere velocities ($U_p$, $V_p$ and $W_p$) were estimated to be 2\%, 1\% and 4\% of the respective $U_\infty$. 

\section{Results and Discussion}
\label{Results}

\subsection{Sphere Translation}
\label{Translation}
\subsubsection{Sphere Trajectories}
\label{wall-normal}

\begin{figure}[t!]
		
        \centering

        {\includegraphics[width=1\textwidth]{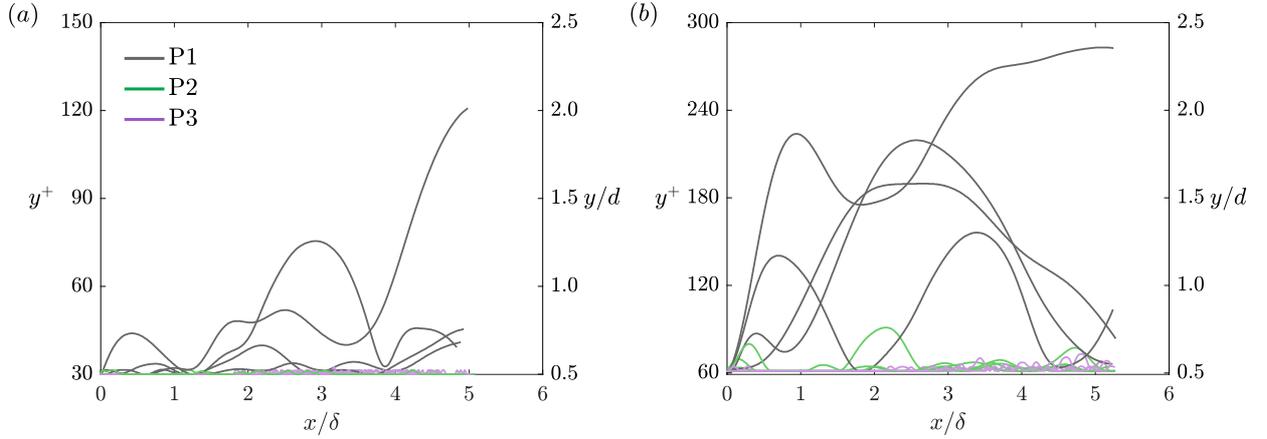}}

	\caption{Samples of ($R=4/10$ runs) sphere wall-normal trajectories at ($a$) $Re_{{\tau}}=680$ and ($b$) $Re_{{\tau}}=1320$ respectively plotted based on the centroid positions. Left axis: plotted in viscous units; Right axis: normalized by the sphere diameter, $d$. Specific gravity: P1 = 1.006 (black), P2 = 1.054 (green) and P3 = 1.152 (purple). 
	}
		\label{fig:2a}
\end{figure}

Figure \ref{fig:2a} shows the sphere wall-normal ($y$) trajectories plotted against their streamwise distance traveled. 
In this section, for visual clarity, only results from four out of ten runs ($R=4/10$) are plotted for each case, unless otherwise specified. 
For results from all ten runs, refer to \cite{tee2019threedimensional}. 
The upward force required to lift a particle off and away from the wall, which must derive from net pressure and viscous forces, must overcome the constant downward buoyancy force, defined as  $F_b = (\rho_p-\rho_f)\pi d^3 g/6$, in all cases studied.
In a turbulent boundary layer, the upward force could derive from mean effects such as mean shear or short-term effects related to local fluid motions.
Depending on the velocity field surrounding the sphere, wall-normal drag can be significant \citep[see e.g.][]{van2013spatially}. 
Also, if $Re_p$ is sufficiently large, i.e. greater than $\sim100$, vortex shedding off of the sphere may generate intermittent wall-normal forces \citep[see][]{zeng2008interactions, van2018experimental}. 
Among these possibilities, we first consider the effect of mean shear on the initial sphere wall-normal motion using \citeapos{hall_1988} equation obtained for a fixed sphere in a turbulent boundary layer such that $\overline{F_y}=F_y^+\nu^2\rho_f$.  If the mean wall-normal fluid-induced force is greater than the net buoyancy force, then $\overline{F_y}^*=(\overline{F_y}-F_b)/F_b>0$, and the sphere is expected to lift off upon release from rest on average (see Table \ref{tab:1}).
The uncertainty of this equation was approximately $\pm0.15\overline{F_y}$ for both flow Reynolds numbers investigated.

For the least dense sphere P1, at $Re_{{\tau}} = 680$, \textcolor{black}{$\overline{F_y}^* = 12$}. 
Increasing $Re_{{\tau}}$ to 1320 doubled $d^+$ so that the estimated mean force increased sixfold.
These estimations correlated very well with our observations for P1 (plotted as black in Fig.\ \ref{fig:2a}).  
At $Re_{{\tau}} = 680$, this sphere lifted off immediately and within approximately $0.3\delta$ or $3d$ in $65\%$ and $18\%$ of 75 runs observed, respectively. 
In the remaining runs, it lifted off at downstream locations significantly greater than $0.3\delta$. 
Meanwhile, at $Re_{{\tau}} = 1320$, this sphere always lifted off immediately upon release in 75 runs observed. 
Owing to the stronger resultant upward force at higher $Re_{{\tau}}$, sphere P1 initially rose to greater heights than at the lower $Re_{{\tau}}$ in nine out of ten  trajectories reconstructed. 
Thus, the initial lift-off height correlated strongly with the local mean shear ($d\overline{U(y)}/dy$) which is 40\% higher for the higher $Re_{{\tau}}$.
In a recent fully-resolved DNS of flow over a sphere initially resting on a porous rough bed, \cite{yousefi2020single} showed a resuspension case with $\overline{F_y}^* = 15$ that lifted off initially up to $y = 2.5d$ in one of the four runs simulated.

Although \textcolor{black}{$\overline{F_y}^* = -0.2$} at $Re_{{\tau}}=680$ for sphere P2, it lifted off of the wall one out of ten runs (plotted as green in Fig.\ \ref{fig:2a}$a$). 
In the remaining runs, this sphere mainly translated along the wall once released.
Meanwhile, at higher $Re_{{\tau}}$ (plotted as green in Fig.\ \ref{fig:2a}$b$), this sphere lifted off of the wall eight out of ten runs due to the stronger mean upward lift force ($\overline{F_y}^* = 3.4$).
At this $Re_{{\tau}}$, all initial lift-off heights were lower than those of P1.

For the densest sphere P3, at the lower $Re_{{\tau}}$, no initial lift-off was observed in 65 runs. 
The sphere did not have sufficient upward lift to overcome the downward buoyance force ($\overline{F_y}^* =  -0.74$) and translated along the wall upon release.
At the higher $Re_{{\tau}}$, although $\overline{F_y}^* =  0.45$, sphere P3 lifted off of the wall only once in 65 runs.

In general, \citeapos{hall_1988} equation provided a good estimate as to whether a sphere lifted off initially.
The variations in behavior observed for sphere P2 (and the one exception observed for P3) are likely explained by instantaneous variations in streamwise velocity approaching and surrounding them.
The r.m.s.\ streamwise fluid velocities at the initial sphere centroid positions at $Re_{{\tau}}=680$ and $1320$ were $20\%$ and $15\%$ of the unperturbed local mean value respectively (see Fig.\ \ref{fig:2}$b$).
Thus, according to the \citeapos{hall_1988} equation, the instantaneous lift forces could vary up to $40\%$ and $26\%$ from $\overline{F_y}$ respectively. 
Similarly, the instantaneous wall-normal drag force can vary depending on the local velocity fields.
These time-dependent variations in the lift and wall-normal drag forces, can also help explain the large variations in initial lift-off heights observed for the lifting spheres released under the same mean fluid condition.
Finally, it is important to point out that \citeapos{hall_1988} equation was fitted based on experimental data up to $Re_p=1250$. 
Hence, our observations on sphere P3 at $Re_{{\tau}}=1320$ with $Re_p=1840$ suggest that \citeapos{hall_1988} equation might over-predict the mean wall-normal fluid-induced force for spheres with $Re_p>1250$. 

Spheres that initially lifted off always descended towards the wall after reaching a local maximum in height. 
While the net buoyancy force was downward in all cases, these descents could be aided by changes in the wall-normal force after the spheres moved away from the wall (see \cite{zeng2008interactions} and additional discussion in Section \ref{Wall-normalV} below).
Sphere P1 at $Re_{{\tau}}=1320$ either recontacted the wall and then lifted off (saltation) or else reached a minimum height above the wall before reascending to a higher location (resuspension).
In all other lifting cases, the spheres always recontacted the wall and then either lifted off again (saltation) or else translated along the wall (sliding and/or rolling). 
These propagation modes were previously observed by \cite{francis1973experiments}, \cite{abbott1977saltation}, \cite{ sumer1981particle}, \cite{ nino1994gravel} and \cite{van2013spatially} with particles propagating over rough or smooth beds.

During either saltation or resuspension, sphere P1 frequently ascended to greater heights than those attained after the initial release, with the maximum height observed of $y\approx2.3d$.  
This behavior was observed in 10/10 and 7/10 runs investigated at $Re_\tau=680$ and $1320$ respectively. 
At $Re_\tau=680$, this sphere collided with the wall and saltated more frequently than at $Re_\tau=1320$ where it mostly propagated above and further away from the wall.
By contrast, the maximum heights observed for spheres P2 and P3 were 0.3$d$ or less throughout their entire trajectories.
Interestingly, sphere P3 exhibited a consistent lift-off pattern that was different from the other spheres.
Although this sphere did not generally lift off upon release, after it had traveled along the wall a certain distance, repeated small lift-off events began to take place (see figures \ref{fig:2b}($c$, $d$) for better illustration). 
This behavior, which was observed for both $Re_\tau$, typically began when $x\gtrsim 2\delta$.
Even though sphere P1 lifted off to a larger height than other spheres throughout the trajectories, its rising and falling angles were always less than $12^\circ$, and typically stayed below $5^\circ$.
For spheres P2 and P3, the rising and falling angles were typically $2-3^\circ$.

\subsubsection{Sphere Streamwise Velocity}
\label{StreamwiseV}

\begin{figure}[t!]
        \centering
       
		\includegraphics[width=.95\textwidth]{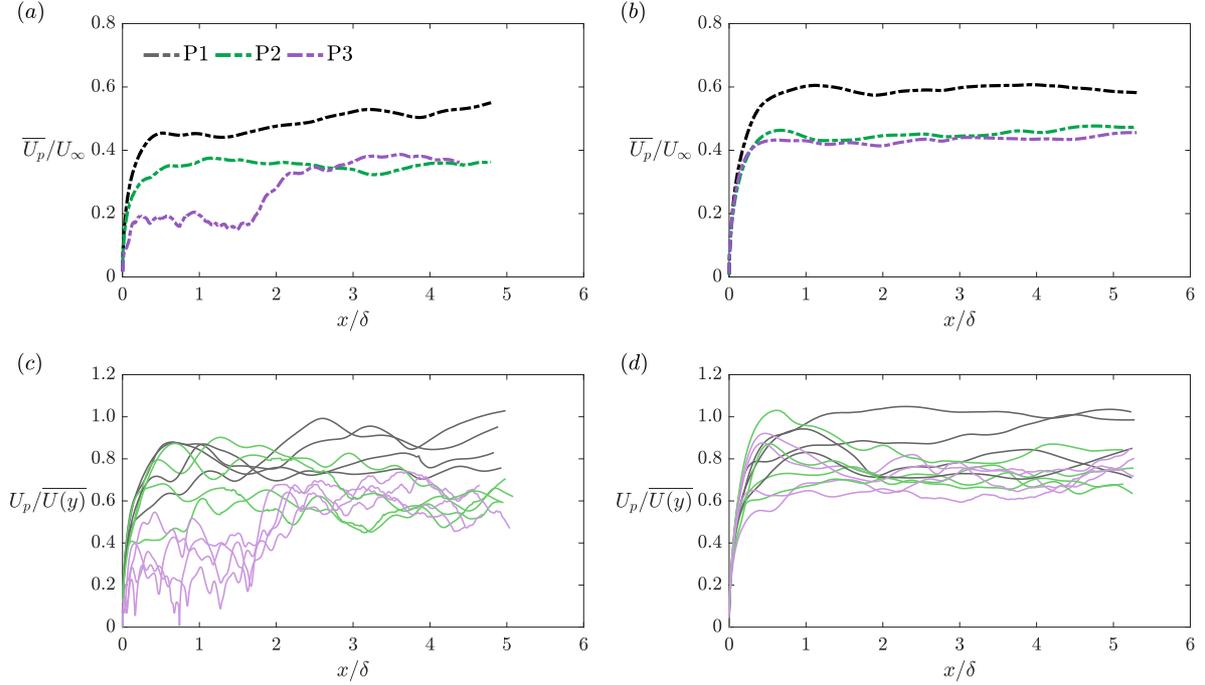}

	\caption{($a$, $b$) Sphere mean streamwise velocity, $\overline{U_p}$ normalized by free stream velocity, $U_\infty$ at $Re_{{\tau}}=680$ and $Re_{{\tau}}=1320$ respectively. ($c$, $d$) Samples of ($R=4/10$ runs) sphere streamwise velocity, $U_p$ normalized by mean unperturbed streamwise fluid velocity ($\overline{U(y)}$) at the height of the sphere centroid at $Re_{{\tau}}=680$ and $Re_{{\tau}}=1320$ respectively. Black: P1; Green: P2; Purple: P3. }
	
	\label{fig:4}
	\vspace{-0.2cm}
\end{figure}

Figures \ref{fig:4}($a$, $b$) represent the sphere mean streamwise velocities ($\overline{U_p}$) obtained based on averaging $U_p$ at each streamwise location over $R=10$ curves for each case.
Once released, the spheres gained momentum from the hydrodynamic drag force and accelerated rapidly with the flow.
At a streamwise location close to $0.5\delta$, the acceleration magnitudes decreased sharply and the spheres began to approach an approximate mean terminal velocity.
From there onward, the mean velocity curves stay relatively flat for all spheres except for sphere P3 at $Re_\tau=680$ where the sphere accelerated again near $x\approx1.5\delta$ to a new approximate mean terminal velocity.
Interestingly, even though the details of sphere motions between P2 and P3 were comparatively different, at both $Re_\tau$, their mean velocity curves tend towards similar values.
Note that in all cases, after achieving approximate terminal velocity ($x>2\delta$), the statistical uncertainty in $\overline{U_p}$, estimated based on $\pm \frac{t\sigma}{\sqrt{R}}$ where $\sigma$ is the standard deviation and $t=2.262$ for $95\%$ confidence level, is always less than $\pm 15\%$ of the respective mean value.

Considering spheres that propagated along the wall, the initial acceleration of sphere P3 at $Re_\tau=680$ was much smaller than for sphere P2 at $Re_\tau=680$ and sphere P3 at $Re_\tau=1320$.
These trends can be interpreted by taking into account the forward drag and opposing wall friction forces acting on each sphere at the time of release.
We estimate the initial drag force using $F_D=\pi\rho_fC_D(U_{rel}d)^2/8$ where $C_D$ is the drag coefficient obtained from the standard drag curve \citep[]{clift2005bubbles} and $U_{rel}$ is the initial relative fluid velocity at the sphere centroid location. 
Then, the friction force, defined as $F_f=f_c N$ where $f_c$ is a constant friction coefficient and $N=F_b-\overline{F_y}$ is the approximate normal force for a non-lifting sphere, is computed for sphere P2 and P3.
At $Re_\tau=680$, as the sphere density increased from P2 to P3, friction increased by an order of magnitude. 
Hence, sphere P3 experienced stronger opposing wall friction and accelerated more weakly than P2 upon release. 
As $Re_{{\tau}}$ increased to 1320, the forward drag increased almost fourfold. Due to the smaller forward drag force sphere P3 experienced at $Re_\tau=680$ than at $Re_\tau=1320$, its forward motion was strongly retarded. Note also that for sphere P3, as $Re_{{\tau}}$ increases, the friction force should decrease due to the concomitant increase in $\overline{F_y}$.

In a uniform, steady, unbounded flow, a sphere would accelerate until it approaches the surrounding fluid velocity. 
However, in our studies, the presence of both turbulence and the wall modify both the surrounding flow fields and the sphere kinematics.
Based on figures \ref{fig:4}($c$, $d$), where $U_p$ is normalized by the mean unperturbed fluid velocity at the height of the sphere centroid ({$\overline{U(y)}$}) obtained from PIV, the sphere velocities mostly lagged behind the local mean fluid velocity over the streamwise distance investigated.
Sphere P1 came closer to the mean local fluid velocity than the other particles, with mean velocity of approximately $0.9 \overline{U(y)}$ at both $Re_{{\tau}}$. 
For spheres P2 and P3, the curves began to level off around $0.6 \overline{U(y)}$ and $0.7 \overline{U(y)}$ at $Re_{{\tau}}$ = 680 and 1320 respectively.
These differences are likely due to the fact that P1 is typically detached from the wall while P2 and P3 are not.
Moreover, on some occasions, sphere P1 at both $Re_{{\tau}}$ and sphere P2 at $Re_{{\tau}}=1320$ traveled faster than the local fluid mean.

\begin{figure}[t!]
        \centering
        {\includegraphics[trim={0 0 0mm 0},clip, width=.98\textwidth]{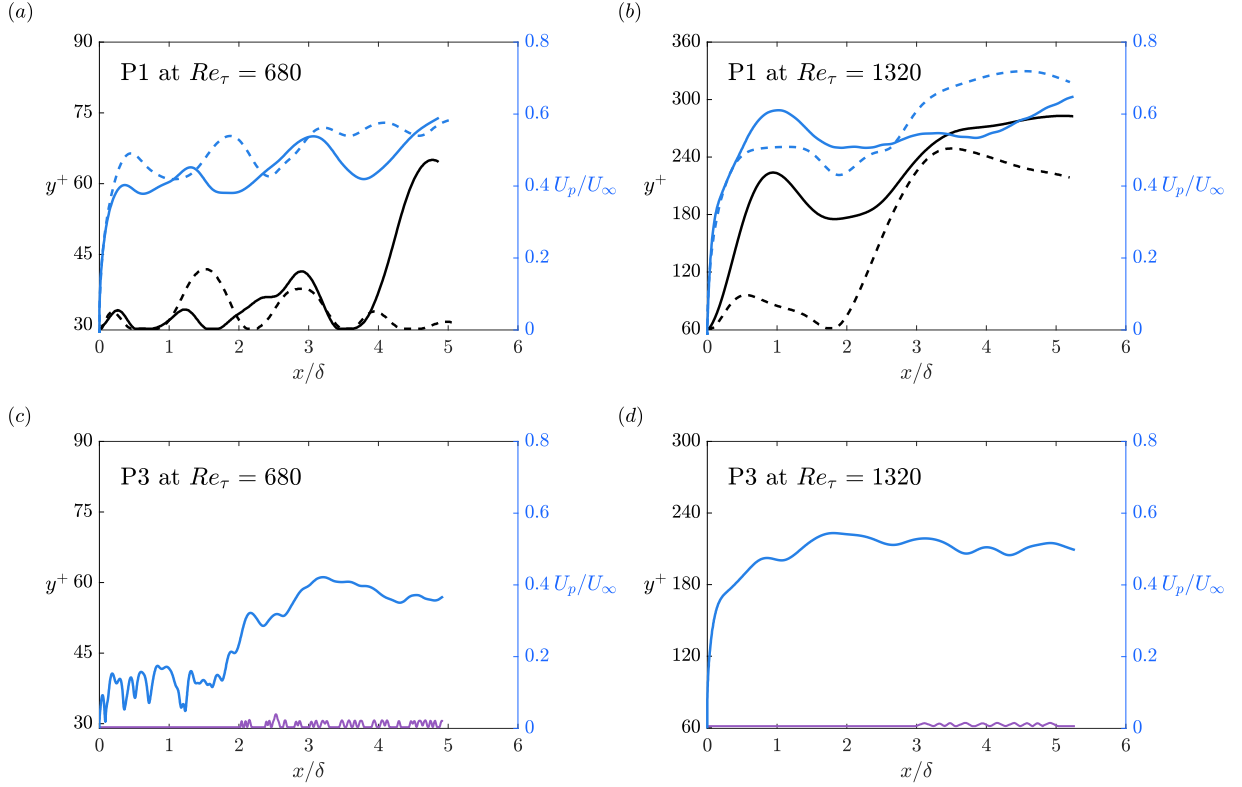}}

	\caption{{Sphere wall-normal trajectories, $y^+$ (left axis) and streamwise velocities, $U_p/U_\infty$ (right axis; blue). ($a$, $c$) $Re_{{\tau}}=680$; ($b$, $d$) $Re_{{\tau}}=1320$. Black: P1; Purple: P3. Solid and dashed lines in ($a$, $b$) represent two individual runs.}}
		\label{fig:2b}
\end{figure}

In all runs, even after attaining an approximate terminal velocity, the individual sphere velocity curves still fluctuate substantially.
To better understand the fluctuations in the velocity curves, sample wall-normal trajectories for sphere P1 (black) and P3 (purple) and their corresponding streamwise velocities (blue) at both $Re_{{\tau}}$ are plotted in Fig.\ \ref{fig:2b}.
For sphere P1, the local $y$-position correlates positively with the respective value of $U_p$ in all four runs shown and in the additional runs corresponding to these cases. 
As sphere P1 ascended, it gained more momentum from the faster moving fluid away from the wall and accelerated; as it descended, it lost momentum due to the slower moving fluid near the wall and thus began to decelerate.
When we estimated the oscillating frequencies ($f$) of the streamwise velocity curves at both $Re_{{\tau}}$, in portions of runs where $Re_p> 100$, the resulting Strouhal number, $St = fd/U_{rel}$ corresponded very well with the range of 0.1 to 0.2 associated with vortex shedding \citep[]{zeng2008interactions, van2018experimental}.
{This behavior was observed frequently for both $Re_{{\tau}}$, and in almost every run at $Re_{{\tau}}=680$ where, downstream of $x=0.5\delta$, $Re_p$ remained larger on average.}

Sphere P3 always stayed close to the wall, and where the height varied during very small lift-off events ($x>2\delta$), it occurs at a much shorter wavelength than the corresponding streamwise velocity fluctuations in the same region (see figures \ref{fig:2b}($c$, $d$)). 
Therefore, these variations appear decoupled.
When $x<2\delta$, the velocity curve for $Re_{{\tau}}=680$ varied strongly at $St\backsim0.2$ also suggesting vortex shedding.
This behavior, which was common across all runs for this case, manifests as short wavelengths in Fig.\ \ref{fig:2b}$c$ due to the relatively small propagation speed and large relative velocity in that region.
After the sphere accelerated again at $x\sim2\delta$, the vortex shedding effect was still present.  
{At $Re_{{\tau}}=1320$, the streamwise velocity curves also fluctuated at the vortex shedding frequency.

While the effect of vortex shedding on sphere streamwise velocity is important in most of the cases studied, the simulations by \cite{zeng2008interactions} on a fixed sphere located above the wall showed that the oncoming wall turbulence was also an important contributor to fluctuations in the streamwise force on the sphere.
Similarly, in some instances, the timewise variations in velocity magnitude observed herein are likely driven by local high and low momentum regions that surround and move past the sphere.
Our preliminary findings on planar fluid velocity fields surrounding a moving sphere at $y=0.75d$ have shown that a sphere enveloped by a long streamwise region of relatively slow moving fluid traveled significantly slower than the same sphere enveloped by relatively fast moving fluid. 
Thus, the local fluid velocity variations can explain the strong variations in sphere velocity at a given streamwise location across different runs as well as some occasions where $U_p/\overline{U(y)}>1$.

\subsubsection{Sphere Wall-normal Velocity}
\label{Wall-normalV}
The wall-normal sphere velocities ($V_p$) at $Re_{{\tau}}=1320$ are plotted in Fig.\ \ref{fig:5a}$a$. 
Two sample $V_p$ curves are also plotted in blue on the right axis in Fig.\ \ref{fig:5a}$b$ along with their respective wall-normal trajectories in black on the left axis.
The results show that upon release, sphere P1 propagated with the highest $V_p$, up to $0.1U_\infty$, and hence lifted off to a greater height than other spheres.  
At the same time, the oscillating frequencies of the wall-normal velocity for P1 in Fig.\ \ref{fig:5a}$b$, as well as in the other runs where $Re_p$ were also $> 100$, matched very well with the Strouhal number associated with vortex shedding.
As sphere P2 has greater density than P1, it lifted off with smaller wall-normal velocity.
{Although the initial $V_p$ of P3 was zero, after $x\approx2\delta$, the curves began to fluctuate at frequencies corresponding with $St\sim O(1)$ or larger, with ascending velocity of approximately $0.05U_\infty$.}
The higher values of $St$ imply that the small repeated lift-off events, which corresponded to the fluctuations in wall-normal velocity, were not prompted by vortex shedding.
Surprisingly, even though sphere P3 was the heaviest sphere and lifted off less than $0.25d$ from the wall, it generally reached $V_p$ magnitudes larger than those of P2 and comparable to those of P1.
Similar wall-normal velocity trends were observed at the lower $Re_{{\tau}}$ for all spheres (not shown here for brevity).

\begin{figure}[t!]
        \centering
        {\includegraphics[width=1\textwidth]{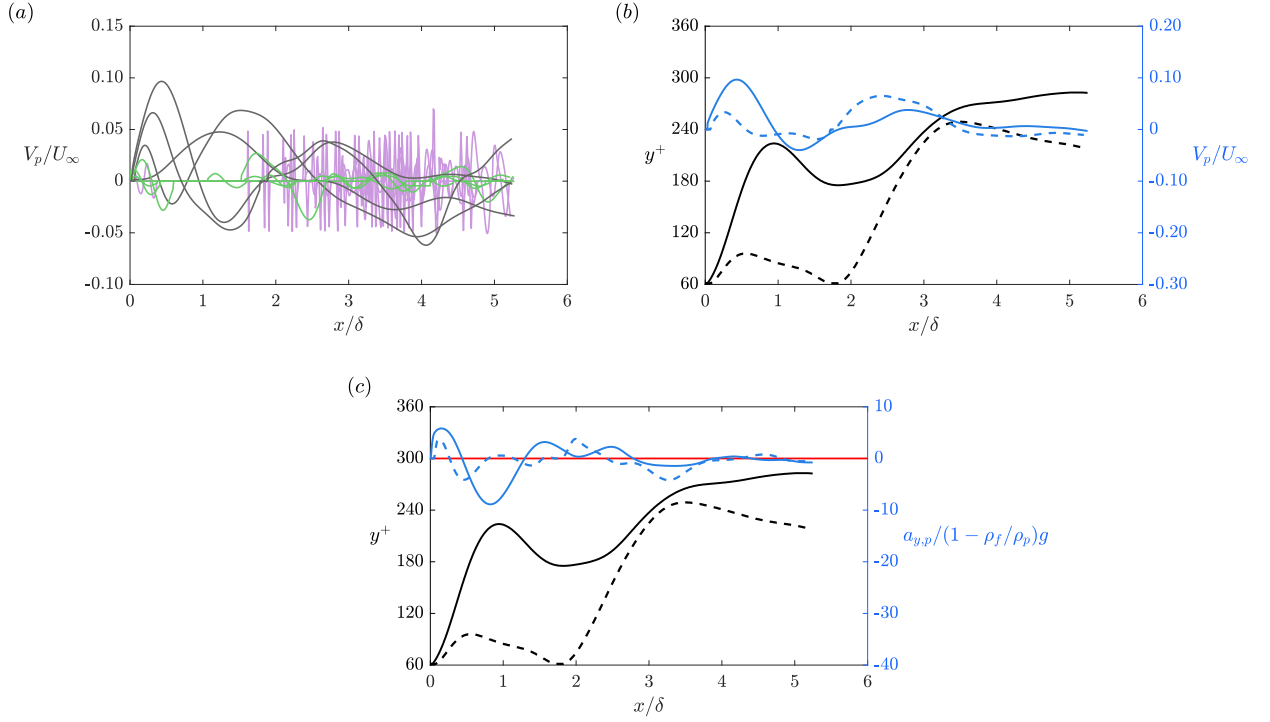}}
      
	\caption{{($a$) Samples of ($R=4/10$ runs) sphere wall-normal velocity at $Re_{{\tau}}=1320$, $V_p$ normalized by free stream velocity, $U_\infty$. Black: P1; Green: P2; Purple: P3. ($b$, $c$) Sphere P1 at $Re_{{\tau}}=1320$; Left axis: wall-normal trajectories, $y^+$ (black); Right axis: sphere wall-normal velocity, $V_p/U_\infty$ and acceleration, $a_{y,p}$ respectively (blue). Solid and dashed lines represent two individual runs plotted. Red line in ($c$) represents $a_{y,p}=0$.}
	}
		\label{fig:5a}
\end{figure}

On the other hand, during descents, as shown in Fig.\ \ref{fig:5a}$a$, P1 moved towards the wall with minimum values close to $V_p\approx-0.06U_\infty$, almost double its settling velocity.
In most runs, P1 descended with negative $V_p$ value larger than or comparable to those observed for both the denser spheres.
A similar finding was observed in the numerical simulation of \cite{yousefi2020single}, where their resuspending sphere also descended with magnitude larger than $V_s$ value at multiple instances along its trajectory.

When the sphere was held at rest in the flow, $Re_p$ was 760 and 1840 for $Re_{{\tau}}=680$ and 1320, respectively.
After strong initial accelerations, these values decreased significantly in all cases. Beyond $x=3\delta$, the average $Re_p$ values were larger for P2 and P3 ($O(300)$ and $O(400)$ at $Re_\tau=680$ and 1320 respectively) than for P1 ($O(200)$ at both $Re_\tau$).
While the initial lift-off events were governed mainly by the mean shear lift \citep[as in][]{hall_1988, yousefi2020single}, the reduction in related upward force associated with the decrease in $Re_p$ should weaken the subsequent lift-off activity.
However, the resulting wall-normal trajectories and velocities of the spheres demonstrated that the sphere lift-off heights could be larger downstream than those upstream. 
This signifies the presence of a strong upward impulse of a similar or larger value than that related to the initial mean shear lift.
Similarly, for instances where the sphere descended faster than its settling velocity, the presence of a downward fluid impulse can be as significant as that due to the negative buoyancy force.

In this context, \cite{zeng2008interactions} inferred that the strong instantaneous positive and negative wall-normal forces observed on a fixed sphere were closely associated with ejection and sweep events. 
Studies by \cite{sutherland1967proposed} and \cite{van2013spatially} on moving particles, for example, have shown that ejection events are responsible for particle lift-offs.
These findings suggest that the subsequent lift-off events of larger magnitudes observed in the current study are likely triggered and aided by coherent fluid motions. 
While ejection events could provide the sphere with upward momentum, negative wall-normal fluid motions such as sweeps could provide the sphere with sufficient downward momentum to descend faster than its settling velocity. 
Depending on the type and strength of coherent structures encountered by a sphere, its wall-normal motion can vary significantly (as shown in Fig.\ \ref{fig:2a}) with the change in instantaneous lift and wall-normal drag forces. 
As a reference, the r.m.s.\ wall normal fluctuating velocity  ($v_{rms}$) for $ 29 < y^+ < 300 $ could vary between 3 and $5\%$ of $U_\infty$ at both $Re_{\tau}$.

It is notable that, over the streamwise distances investigated, all sphere trajectories were limited approximately to the buffer and logarithmic regions where turbulent activity is strongest.  
Thus, although a sphere could be lifted to significant heights above the wall, the wall-normal impulses were not sufficient to drive it to even greater heights.
Considering the nature of the impulse, it can derive from a large upward force exerted over a short time or a smaller force exerted over a longer time. 
For all three spheres examined, the upward impulse must counteract a downward one due to gravity.  
The wall-normal acceleration for sphere P1 plotted in Fig.\ \ref{fig:5a}$c$ revealed multiple behaviors. 
In one run (solid curves), the sphere acceleration, and net impulse, was initially positive until it reached $y^+ = 150$. 
Further downstream, beginning at $x=1.3\delta$, the sphere again experienced a sustained upward acceleration over the range $175 < y^+ < 220$.
By contrast, in the other run (dashed curves), the initial upward acceleration lasted only for a short distance until the sphere rose to $y^+ = 80$.  
Further downstream, however, starting at $x=1.8\delta$, the same sphere experienced another upward acceleration as it rose from the wall up to $y^+ = 140$.
The fact that the turbulent activity, and the probability of sustained upward fluid motions, dies off beyond the logarithmic region, helps explain the observed result that the spheres examined never rose above a limit of $y^+ = 282$ ($y/\delta=0.21$) within the field of view investigated.

The wall-normal trajectories plotted in Fig.\ \ref{fig:2a} showed that in most runs, spheres that initially lifted off descended and collided with the wall before lifting off again. 
Thus, to better understand the effect of collision on sphere lift-off events, the coefficient of restitution ($e$), which is the ratio of wall-normal velocity after impact ($V_f$) to wall-normal velocity before impact ($V_i$), was computed. 
Among all collision incidents, $e$ was always zero. 
After colliding with the wall, the spheres slid for a distance of at least $0.1d$ ($\geqslant3$ frames) before lifting off again.
When the impact Stokes numbers, defined as \textcolor{black}{$Sk = \rho_pV_id/18\nu\rho_f$}, are calculated, in all trajectories, {$Sk<10$}. 
Therefore, the $e=0$ result for \textcolor{black}{$Sk<10$} is consistent with previous experiments on particles moving through quiescent fluid and impacting on  solid surfaces \citep[]{joseph2001particle, gondret2002bouncing}.

\subsubsection{Sphere Spanwise Motion and Velocity}
\label{spanwise}
Sphere trajectories in the streamwise-spanwise plane are plotted in Fig.\ \ref{fig:3}.
Once released, instead of propagating along $z=0$, the spheres typically moved sideways, occasionally reversing direction, with some crossings over $z=0$.
For P3 at $Re_{{\tau}}=680$ and $x<2\delta$, the spanwise trajectories exhibited shorter wavelength fluctuations than the other cases.
In all cases, the maximum spanwise distance traveled from $z=0$ was approximately $12\%$ of the corresponding streamwise distance traveled.
This magnitude is similar to that reported by \cite{yousefi2020single} for their resuspension case where the sphere traveled mostly away from the bed.
Considering all cases, including the lifting sphere P1 at $Re_{{\tau}}=1320$, in most runs, the absolute spanwise migration magnitudes were comparable to or larger than the lift-off magnitudes (see figures \ref{fig:5a}$b$ and \ref{fig:5b}$b$).

\begin{figure}[t!]

        \centering
    			 \includegraphics[width=.98\textwidth]{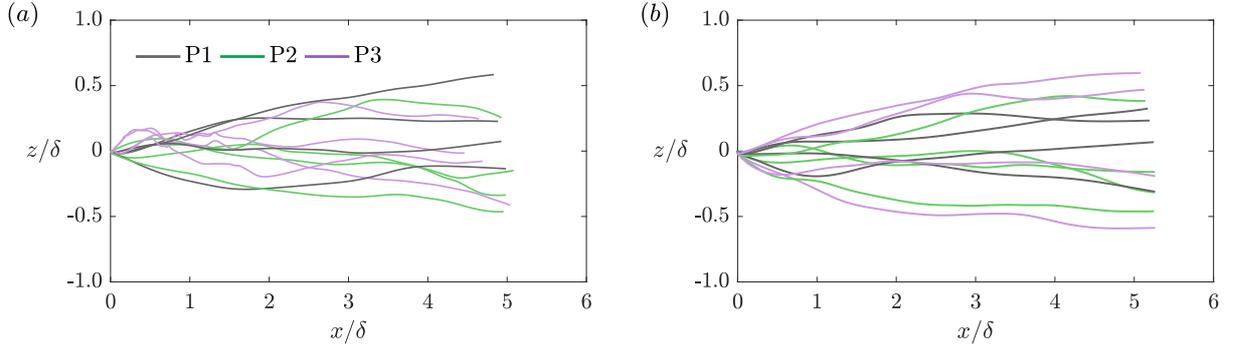}

	 		\caption{Samples of ($R=4/10$ runs) sphere spanwise trajectories, $z$ at ($a$) $Re_{{\tau}}=680$ and ($b$) $Re_{{\tau}}=1320$ respectively.}
        \label{fig:3}
\end{figure}

\begin{figure}[t!]
        \centering
        {\includegraphics[width=1\textwidth]{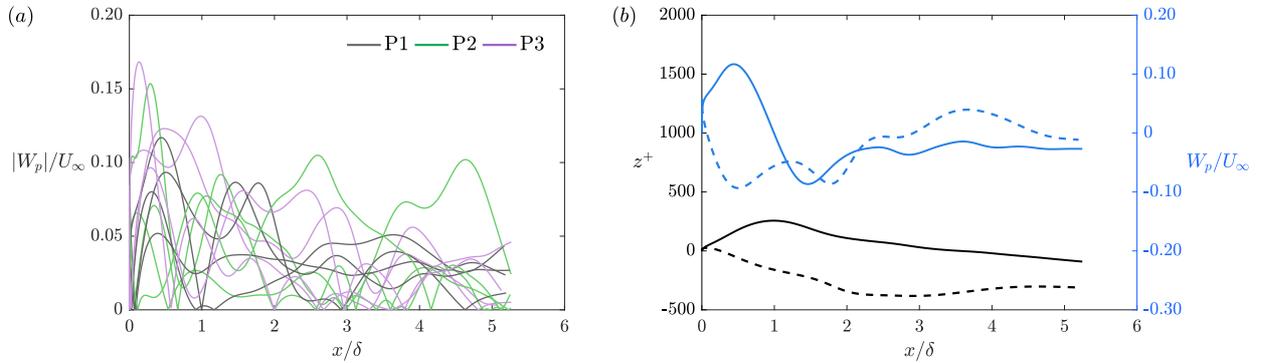}}
      
	\caption{($a$) Samples of ($R=4/10$ runs) sphere absolute spanwise velocities, $|W_p|$ normalized by free stream velocity. Black: P1; Green: P2; Purple: P3. ($b$) Sphere P1 at $Re_{{\tau}}=1320$. Left axis: spanwise trajectories, $z^+$ (black); Right axis: spanwise velocities, $W_p$ (blue). Solid and dashed lines correspond to the two individual runs plotted in Fig.\ \ref{fig:5a}$b$. 
	}
		\label{fig:5b}
\end{figure}

The absolute spanwise velocities ($|W_p|$) for spheres at $Re_{{\tau}}=1320$ are illustrated in Fig.\ \ref{fig:5b}$a$.
Overall, $|W_p|$ varied over larger magnitudes than $|V_p|$.
By comparing $W_p$ with $V_p$ for sphere P1 at $Re_{{\tau}}=1320$ (see sample runs in figures \ref{fig:5a}$b$ and \ref{fig:5b}$b$), including during the lift-off events, the magnitude of $W_p$ typically exceeded that of $V_p$. 
At the same time, the fluctuations in $W_p$ in both runs shown corresponded directly with those appearing in $V_p$ at the vortex shedding frequency. 
{This correspondence was apparent as well in portion of other runs where $Re_p>100$ as highlighted in Section \ref{StreamwiseV}}.
Despite the distinct differences in $V_p$ curves for spheres of different specific gravities, P2 and P3 attained values of $W_p$ comparable to those of P1 throughout their trajectories.
Even though the wall-normal and spanwise velocity curves appeared uncorrelated for both P2 and P3, the spanwise velocity curves also fluctuated at $St \sim 0.1$-0.2 suggesting vortex shedding in those cases as well.
Similar spanwise velocity trends were observed at the lower $Re_{{\tau}}$ {for all spheres} (not shown here for brevity).
For comparison, at $Re_\tau=1320$ and $29<y^+<300$, the fluid spanwise r.m.s.\ exceeds the wall-normal r.m.s.\ value, e.g.,  at $y^+=60$, $w_{rms}/U_\infty=0.06$ and $v_{rms}/U_\infty=0.04$ \citep[]{jimenez2010turbulent}.
Therefore, within the buffer and logarithmic regions, spanwise forces exerted by the fluid can be larger than the comparable wall-normal forces \citep[see also][]{zeng2008interactions}.
Our preliminary results including fluid velocity fields reveal that the sphere spanwise motion is significantly affected by spanwise fluid drag.
For example, we observed that sphere enveloped by the low momentum region tended to travel in the spanwise direction together with the surrounding fluid.  
More quantitative findings on the correlation between sphere and fluid spanwise velocities are under investigation.

\subsection{Sphere Rotation}
\label{Rotation}

Figure \ref{fig:6} depicts sphere orientation components for $R=4/10$ runs of all six cases investigated.
The orientations were calculated by integrating the respective angular velocities obtained from the rotation matrix.
Depending on the mode of translation, two distinct trends were observed.
Spheres that mostly traveled above the wall, namely sphere P1 at both $Re_{{\tau}}$ and sphere P2 at $Re_{{\tau}}=1320$, rotated less than half a revolution about all axes throughout the range examined.
These weak rotations occurred mostly during the acceleration stage when $x<0.5\delta$.
{Notably, even in the presence of strong initial mean shear, the spheres did not develop any significant forward rolling motion (rotation about the negative $z$-axis).}
Also, no significant correlations were found between rotation about the streamwise axis ($\theta_x$) and spanwise migration. 
Without wall friction, these small rotations were most likely induced by the flow structures such as individual vortices or shearing regions.

\begin{figure}[t!]
        \centering
        		\includegraphics[width=1\textwidth]{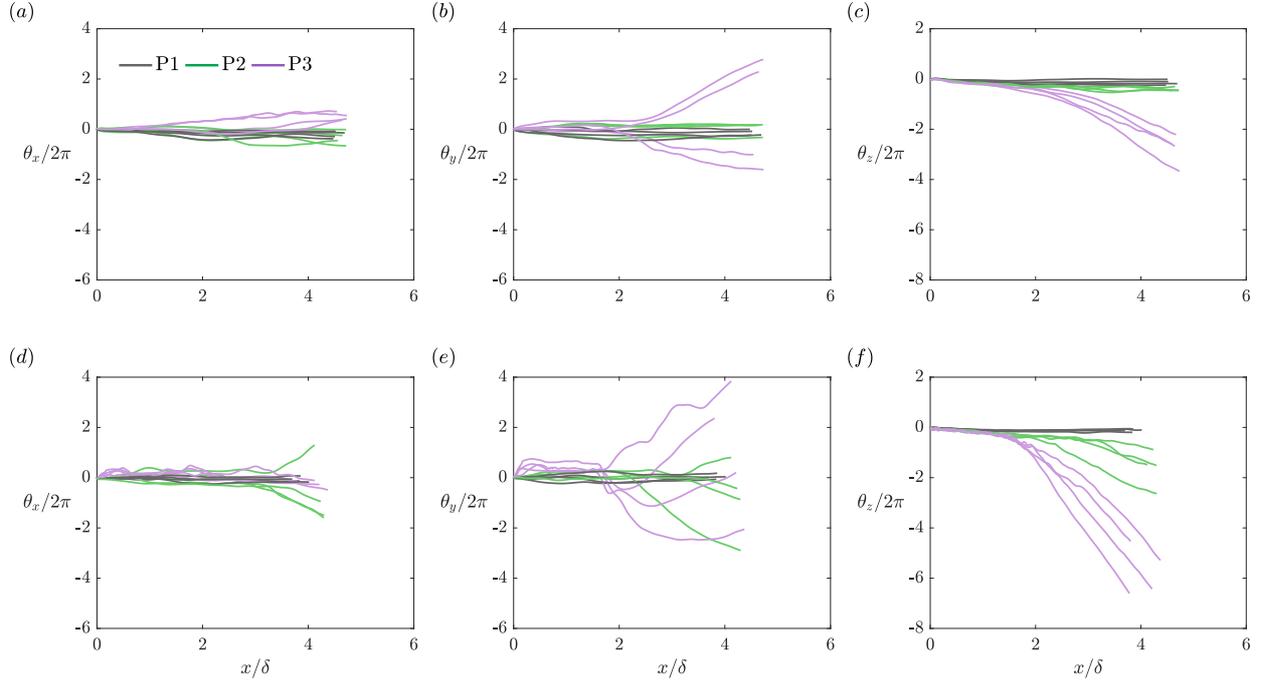}
        
	\caption{Samples of ($R=4/10$ runs) sphere orientation tracks (${\theta_x}$, ${\theta_y}$ and ${\theta_z}$) at $Re_{{\tau}} = 1320$ ($a$-$c$) and $Re_{{\tau}} = 680$ ($d$-$f$) , respectively.} 
	\label{fig:6}

\end{figure}

By contrast, spheres that mostly traveled along the wall, namely sphere P2 at $Re_{{\tau}}=680$ and sphere P3 at both $Re_{{\tau}}$, developed significant rotations starting from $x\gtrsim1.5\delta$.
Upon release, these spheres mainly slid along the wall and barely rolled.
For $x\lesssim1.5\delta$, their $\theta_z$ magnitudes were smaller than either $\theta_x$ or $\theta_y$, and of similar magnitude to those of the lifting spheres.
Further downstream, however, these spheres began to roll forward.
This is clearly indicated by the sharp and steady decrease of $\theta_z$ values in the plots.
While rolling forward, these spheres also rotated about the $y$-axis significantly, exhibiting coupled rotation behavior.
This wall-normal rotation could possibly be triggered by adjacent fast and slow moving zones in the boundary layer which would generate a hydrodynamic torque about the $y$-axis.

\begin{figure}[t!]
        \centering
        		\includegraphics[trim={1mm 0 0mm 0},clip,width=1\textwidth]{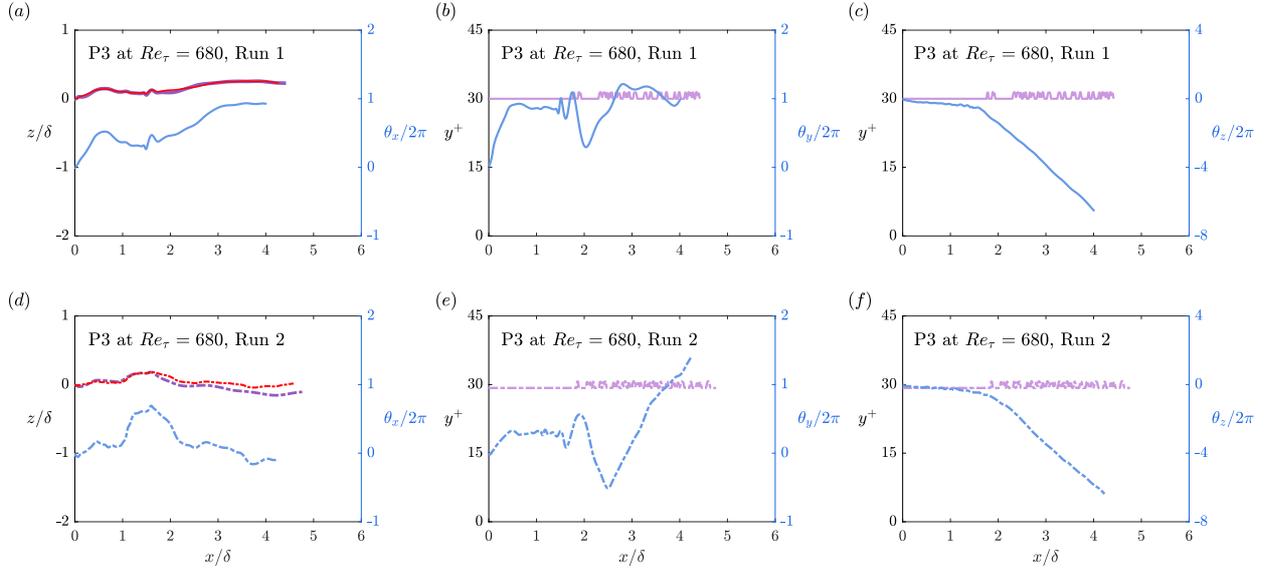}
        
	\caption{Sphere P3 at $Re_{{\tau}} = 680$. Solid and dashed lines represent two individual runs. ($a$, $d$) Left axis: sphere spanwise trajectories, $z/\delta$ (purple) and angular displacement, $s_z$ (red); Right axis: sphere rotational angle about streamwise axis, $\theta_x$ (blue). ($b$, $e$) Left axis: sphere wall-normal trajectories, $y^+$ (purple); Right axis: sphere rotational angle about wall-normal axis, $\theta_y$ (blue). ($c$, $f$) Left axis: $y^+$ (purple); Right axis: sphere rotational angle about spanwise axis, $\theta_z$ (blue).} 
	\label{rot}

\end{figure}

Within the field of view investigated, for the wall-interacting spheres, $\theta_x$ magnitudes were the smallest when compared to $\theta_y$ and $\theta_z$.
For spheres P2 and P3 at $Re_{{\tau}}=680$, the fluctuations in $\theta_x$ curves correlated very well with the spanwise trajectories in most runs.
In figures \ref{rot} ($a$, $d$), two examples of spanwise trajectories for sphere P3 at $Re_{{\tau}}=680$ are plotted as purple on the left axis, while their orientations about the streamwise axis are plotted as blue on the right axis.
In both examples, when $z$ is positive, $\theta_x$ is positive and when $z$ is negative, $\theta_x$ is negative.
Moreover, the spanwise displacements calculated based on $s_z = \theta_xd/2$ for a pure rolling motion (plotted as red relative to the left axis) matched very well with the $z$-trajectories as shown in figures \ref{rot} ($a$, $d$).
This implies that the spanwise sphere motions were induced by the streamwise rotations, and while propagating sideways, these spheres were mostly rolling about the $x$-axis instead of sliding.
These rotations could possibly be initiated by quasi-streamwise vortices or spanwise motions in the fluid.

To quantify the relative importance of sliding and forward rolling, we define a dimensionless rotation rate, $\alpha_z = |\Omega_z|d/2U_p$, which relates the sphere rotational velocity about the spanwise axis, $\Omega_z$ to its streamwise translational velocity.
If $\alpha_z$ is 0, the sphere is undergoing pure translation or sliding in the streamwise direction; if $\alpha_z$ is 1 and the sphere is in contact with the wall, the sphere is undergoing pure forward rolling without slipping. 
The dimensionless rotation rates were calculated based on the slope of the mean $x$-trajectory and $\theta_z$ curves for the wall-interacting spheres. 
For the densest sphere P3, when $x>3\delta$, the mean $\alpha_z$ values at $Re_{{\tau}}=680$ and 1320 were approximately 0.6 and 0.4 respectively.
This implies a stronger rolling-to-sliding tendency at the downstream location as $Re_{{\tau}}$ decreases.
Although the $\theta_z$ slope of sphere P2 at $Re_{{\tau}}=680$ also increased over time, the mean value of $\alpha_z \sim0.2$ infers that translation and sliding remained more important than rotation within the range investigated.
Overall, sphere P3 at $Re_{{\tau}}=680$ exhibited the strongest rotations.
As discussed in Section \ref{StreamwiseV}, the denser sphere P3 experienced higher wall friction than P2.
Hence, at $Re_{{\tau}}=680$, it rolled forward with larger angular velocity than P2.
Also, as $Re_{{\tau}}$ increased to 1320, the wall friction weakened and so did the forward rotation.

Finally, we discuss the forward rolling and small repeated lift-off events observed with sphere P3. 
At $Re_{{\tau}}=680$, we noticed that prior to lifting off at $x\approx2\delta$, the sphere had already begun to roll forward and also rotate about both $x$- and $y$-axes (see Fig.\ \ref{rot}).
After the sphere began rolling, the streamwise velocity also increased steeply (see figures \ref{fig:4} and \ref{fig:2b}$c$).
It is unclear how the rolling initiates. 
In the sliding P3 case, gravity dominates the wall normal force so that torque from wall friction is relatively constant along the trajectory.
On the other hand, the forward drag force decreases significantly with increasing sphere velocity, and any torque from fluid friction is also likely to change on average.
Further, time-wise variations in the approaching fluid velocity field may cause significant variations in torque.
Once the sphere begins rolling, Magnus lift can play a role.
\citeapos{kim2014inverse} experiment on spinning spheres in uniform flow reported that as $\alpha_z$ increased from 0.1 to 0.6, $C_L$ increased from approximately 0.05 to 0.3.
These values are comparable to the initial $C_L$ estimated from \citeapos{hall_1988} equation where $C_L=8\overline{F_y}/\pi\rho_f(U_{rel}d)^2 \approx 0.2$.
Therefore, the contribution of a Magnus force in the P3 lift-off events is likely to be important. 
It is also notable that the weaker rotations of P2 appeared insufficient to generate these high frequency lift-offs. 
For P3, the Magnus lift and repeated lift-offs must reduce wall friction and contribute to the second acceleration stage observed.

\section{Conclusions}
\label{Section4}

Three-dimensional translation and rotation of spheres with diameter of 58 and 122 viscous units at $Re_{{\tau}}=680$ and 1320 were reconstructed successfully for multiple cases where the relative importance of gravity and mean shear was varied.
Particles released from rest were tracked individually over a streamwise distance up to $x\approx5.5\delta$. 
Among all cases, two distinct types of dynamics were observed based on the sphere wall-normal trajectories and rotation behaviors. 
They included lifting spheres (sphere P1 at both $Re_{{\tau}}$ and sphere P2 at $Re_{{\tau}}=1320$) and wall-interacting spheres (sphere P2 at $Re_{{\tau}}=680$ and sphere P3 at both $Re_{{\tau}}$).

Upon release, the lifting spheres separated from the wall in most runs when the mean shear lift force was larger than the net buoyancy force. 
Due to velocity variations within the turbulent boundary layer, the initial fluid-induced force and lift-off height varied.
Since the spheres always reached a maximum height in the logarithmic region before descending toward the wall, it is clear that any initial upward force, possibly induced by mean shear lift, upward moving fluid, or even vortex shedding, decreased significantly as the sphere suspended.
After the period of descent, the spheres either ascended again without returning to the wall or else contacted the wall and then slid before lifting off again to heights reaching $y\approx2.3d$ (or $y^+\approx135$ and 282 at $Re_\tau=680$ and 1320 respectively) for the least dense sphere.
Although these spheres mostly traveled above the wall, they nevertheless lagged behind the mean fluid velocity even after attaining an approximate average terminal velocity, propagating with $Re_p\sim O(200)$.
Hence, vortex shedding could typically be present, affecting both the surrounding flow fields and the sphere motions.
Since the mean shear lift decreases with $Re_p$, all subsequent lift-off events, especially those that reached heights larger than the initial maximum values, are likely prompted by local fluid upwash or local variations in the velocity field, including those related to vortex shedding when $Re_p>100$.
Sometimes the fluid motions acted in the opposite direction, e.g., pushing sphere P1 downward with wall-normal velocity double its gravitational settling velocity.
Throughout their propagation, the lifting spheres translated with very weak or minimal rotations about any axis.
 
Upon release, the wall-interacting spheres first slid along the wall with minimal rotation. 
At $Re_{{\tau}}=680$, the initial acceleration of the dense sphere P3 was significantly retarded by the opposing friction force, in contrast to the other spheres that accelerated steeply over a streamwise distance of $\delta$.
In this region, vortex shedding led to strong and regular variations in the streamwise velocity.
After the wall-interacting spheres propagated downstream by $\approx1.5\delta$, forward rolling (with slipping) followed by repeated small lift-off events of magnitude less than $0.1d$ began to occur. 
Since the collisions were completely inelastic ($e=0$), and the wall-normal velocity fluctuated at a frequency higher than that expected for vortex shedding, the repeated lift-off events were not caused either by rebounding or vortex shedding.
Instead, Magnus lift is most likely the reason behind the lift-off events.
While rolling forward, the spheres also rotated about the $x$- and $y$-axes, exhibiting coupled rotation behavior.

In all cases, the spheres moved sideways significantly.
Due to direct wall interactions, the spanwise motions of the denser spheres were strongly correlated with rotation about the $x$-axis.
Even though the lifting spheres mostly traveled above the wall, they still migrated up to 12\% of the streamwise distance traveled.  
Their spanwise velocities were mostly equal to or larger than the wall-normal velocities and fluctuated at the vortex shedding frequency. 
Hence, spanwise forces were substantial in all cases and could be prompted either by rolling about the $x$-axis, shed vortices, or spanwise fluid motions.

To summarize, the kinematics of spheres in a boundary layer are complicated.
The coherent structures in the boundary layer such as high and low momentum regions, as well as vertical and spanwise fluid motions, can have significant effects on both translation and rotation.
Since the finite-size spheres propagated toward their own wakes with velocities smaller than the local mean flow, vortex shedding effects were likely significant in all cases investigated such that some of the sphere velocity variations corresponded with expected shedding frequencies.
Wall friction was important in impeding acceleration of the denser spheres as well as in prompting the rolling motions. 
Recently, we have conducted simultaneous stereoscopic PIV and sphere tracking experiments to resolve fluid motions surrounding a moving sphere. 
These results will be analyzed and discussed in a future communication to complement the current observations and better understand both the particle-wall and particle-turbulence interactions.

\section*{Acknowledgments}
\noindent{The authors thank Nicholas Morse, Ben Hiltbrand and Alessio Gardi for their help with this project. This work was funded by the U.S.\ National Science Foundation (NSF CBET-1510154)}.

\bibliography{References}

\begin{thebibliography}{62}
\expandafter\ifx\csname natexlab\endcsname\relax\def\natexlab#1{#1}\fi
\providecommand{\url}[1]{\texttt{#1}}
\providecommand{\href}[2]{#2}
\providecommand{\path}[1]{#1}
\providecommand{\DOIprefix}{doi:}
\providecommand{\ArXivprefix}{arXiv:}
\providecommand{\URLprefix}{URL: }
\providecommand{\Pubmedprefix}{pmid:}
\providecommand{\doi}[1]{\href{http://dx.doi.org/#1}{\path{#1}}}
\providecommand{\Pubmed}[1]{\href{pmid:#1}{\path{#1}}}
\providecommand{\bibinfo}[2]{#2}
\ifx\xfnm\relax \def\xfnm[#1]{\unskip,\space#1}\fi
\bibitem[{Abbott and Francis(1977)}]{abbott1977saltation}
\bibinfo{author}{Abbott, J.E.}, \bibinfo{author}{Francis, J.R.D.},
  \bibinfo{year}{1977}.
\newblock \bibinfo{title}{Saltation and suspension trajectories of solid grains
  in a water stream}.
\newblock \bibinfo{journal}{Philosophical Transactions of the Royal Society of
  London. Series A, Mathematical and Physical Sciences} \bibinfo{volume}{284},
  \bibinfo{pages}{225--254}.
\bibitem[{Ardekani and Brandt(2019)}]{ardekani2019turbulence}
\bibinfo{author}{Ardekani, M.N.}, \bibinfo{author}{Brandt, L.},
  \bibinfo{year}{2019}.
\newblock \bibinfo{title}{Turbulence modulation in channel flow of finite-size
  spheroidal particles}.
\newblock \bibinfo{journal}{Journal of Fluid Mechanics} \bibinfo{volume}{859},
  \bibinfo{pages}{887--901}.
\bibitem[{Auton(1987)}]{auton1987lift}
\bibinfo{author}{Auton, T.R.}, \bibinfo{year}{1987}.
\newblock \bibinfo{title}{The lift force on a spherical body in a rotational
  flow}.
\newblock \bibinfo{journal}{Journal of Fluid Mechanics} \bibinfo{volume}{183},
  \bibinfo{pages}{199--218}.
\bibitem[{Bagchi and Balachandar(2002)}]{bagchi2002effect}
\bibinfo{author}{Bagchi, P.}, \bibinfo{author}{Balachandar, S.},
  \bibinfo{year}{2002}.
\newblock \bibinfo{title}{Effect of free rotation on the motion of a solid
  sphere in linear shear flow at moderate {R}e}.
\newblock \bibinfo{journal}{Physics of Fluids} \bibinfo{volume}{14},
  \bibinfo{pages}{2719--2737}.
\bibitem[{Balachandar and Eaton(2010)}]{balachandar2010turbulent}
\bibinfo{author}{Balachandar, S.}, \bibinfo{author}{Eaton, J.K.},
  \bibinfo{year}{2010}.
\newblock \bibinfo{title}{Turbulent dispersed multiphase flow}.
\newblock \bibinfo{journal}{Annual Review of Fluid Mechanics}
  \bibinfo{volume}{42}, \bibinfo{pages}{111--133}.
\bibitem[{Barros et~al.(2018)Barros, Hiltbrand and
  Longmire}]{barros2018measurement}
\bibinfo{author}{Barros, D.}, \bibinfo{author}{Hiltbrand, B.},
  \bibinfo{author}{Longmire, E.K.}, \bibinfo{year}{2018}.
\newblock \bibinfo{title}{Measurement of the translation and rotation of a
  sphere in fluid flow}.
\newblock \bibinfo{journal}{Experiments in Fluids} \bibinfo{volume}{59},
  \bibinfo{pages}{104}.
\bibitem[{Bluemink et~al.(2008)Bluemink, Lohse, Prosperetti and van
  Wijngaarden}]{bluemink2008sphere}
\bibinfo{author}{Bluemink, J.J.}, \bibinfo{author}{Lohse, D.},
  \bibinfo{author}{Prosperetti, A.}, \bibinfo{author}{van Wijngaarden, L.},
  \bibinfo{year}{2008}.
\newblock \bibinfo{title}{A sphere in a uniformly rotating or shearing flow}.
\newblock \bibinfo{journal}{Journal of Fluid Mechanics} \bibinfo{volume}{600},
  \bibinfo{pages}{201--233}.
\bibitem[{Casta{\~n}eda et~al.(2014)Casta{\~n}eda, Avlijas, Simard and
  Ricciardi}]{castaneda2014microplastic}
\bibinfo{author}{Casta{\~n}eda, R.A.}, \bibinfo{author}{Avlijas, S.},
  \bibinfo{author}{Simard, M.A.}, \bibinfo{author}{Ricciardi, A.},
  \bibinfo{year}{2014}.
\newblock \bibinfo{title}{Microplastic pollution in {S}t. {L}awrence river
  sediments}.
\newblock \bibinfo{journal}{Canadian Journal of Fisheries and Aquatic Sciences}
  \bibinfo{volume}{71}, \bibinfo{pages}{1767--1771}.
\bibitem[{Cherukat et~al.(1999)Cherukat, McLaughlin and
  Dandy}]{cherukat1999computational}
\bibinfo{author}{Cherukat, P.}, \bibinfo{author}{McLaughlin, J.B.},
  \bibinfo{author}{Dandy, D.S.}, \bibinfo{year}{1999}.
\newblock \bibinfo{title}{A computational study of the inertial lift on a
  sphere in a linear shear flow field}.
\newblock \bibinfo{journal}{International Journal of Multiphase Flow}
  \bibinfo{volume}{25}, \bibinfo{pages}{15--33}.
\bibitem[{Clauser(1956)}]{clauser1956turbulent}
\bibinfo{author}{Clauser, F.H.}, \bibinfo{year}{1956}.
\newblock \bibinfo{title}{The turbulent boundary layer}, in:
  \bibinfo{booktitle}{Advances in Applied Mechanics}.
  \bibinfo{publisher}{Elsevier}. volume~\bibinfo{volume}{4}, pp.
  \bibinfo{pages}{1--51}.
\bibitem[{Clift et~al.(1978)Clift, Grace and Weber}]{clift2005bubbles}
\bibinfo{author}{Clift, R.}, \bibinfo{author}{Grace, J.R.},
  \bibinfo{author}{Weber, M.E.}, \bibinfo{year}{1978}.
\newblock \bibinfo{title}{Bubbles, drops, and particles}.
\newblock \bibinfo{publisher}{Academic Press, Inc., New York}.
\bibitem[{Crowe(2005)}]{crowe2005multiphase}
\bibinfo{author}{Crowe, C.T.}, \bibinfo{year}{2005}.
\newblock \bibinfo{title}{Multiphase flow handbook}.
\newblock \bibinfo{publisher}{CRC press}.
\bibitem[{Dorgan and Loth(2004)}]{dorgan2004simulation}
\bibinfo{author}{Dorgan, A.J.}, \bibinfo{author}{Loth, E.},
  \bibinfo{year}{2004}.
\newblock \bibinfo{title}{Simulation of particles released near the wall in a
  turbulent boundary layer}.
\newblock \bibinfo{journal}{International Journal of Multiphase Flow}
  \bibinfo{volume}{30}, \bibinfo{pages}{649--673}.
\bibitem[{Ebrahimian et~al.(2019)Ebrahimian, Sanders and
  Ghaemi}]{ebrahimian2019dynamics}
\bibinfo{author}{Ebrahimian, M.}, \bibinfo{author}{Sanders, R.S.},
  \bibinfo{author}{Ghaemi, S.}, \bibinfo{year}{2019}.
\newblock \bibinfo{title}{Dynamics and wall collision of inertial particles in
  a solid-liquid turbulent channel flow}.
\newblock \bibinfo{journal}{Journal of Fluid Mechanics} \bibinfo{volume}{881},
  \bibinfo{pages}{872--905}.
\bibitem[{Epps et~al.(2010)Epps, Truscott and Techet}]{epps2010evaluating}
\bibinfo{author}{Epps, B.P.}, \bibinfo{author}{Truscott, T.T.},
  \bibinfo{author}{Techet, A.H.}, \bibinfo{year}{2010}.
\newblock \bibinfo{title}{Evaluating derivatives of experimental data using
  smoothing splines}, in: \bibinfo{booktitle}{Proceedings of Mathematical
  Methods in Engineering International Symposium. MMEI, Lisbon Portugal}, pp.
  \bibinfo{pages}{29--38}.
\bibitem[{Fincham and Spedding(1997)}]{fincham1997low}
\bibinfo{author}{Fincham, A.M.}, \bibinfo{author}{Spedding, G.R.},
  \bibinfo{year}{1997}.
\newblock \bibinfo{title}{Low cost, high resolution {DPIV} for measurement of
  turbulent fluid flow}.
\newblock \bibinfo{journal}{Experiments in Fluids} \bibinfo{volume}{23},
  \bibinfo{pages}{449--462}.
\bibitem[{Fornari et~al.(2016)Fornari, Formenti, Picano and
  Brandt}]{fornari2016effect}
\bibinfo{author}{Fornari, W.}, \bibinfo{author}{Formenti, A.},
  \bibinfo{author}{Picano, F.}, \bibinfo{author}{Brandt, L.},
  \bibinfo{year}{2016}.
\newblock \bibinfo{title}{The effect of particle density in turbulent channel
  flow laden with finite size particles in semi-dilute conditions}.
\newblock \bibinfo{journal}{Physics of Fluids} \bibinfo{volume}{28},
  \bibinfo{pages}{033301}.
\bibitem[{Francis(1973)}]{francis1973experiments}
\bibinfo{author}{Francis, J.R.D.}, \bibinfo{year}{1973}.
\newblock \bibinfo{title}{Experiments on the motion of solitary grains along
  the bed of a water-stream}, in: \bibinfo{booktitle}{Proceedings of the Royal
  Society of London. Series A, Mathematical and Physical Sciences},
  \bibinfo{publisher}{The Royal Society London}. pp. \bibinfo{pages}{443--471}.
\bibitem[{Gao(2011)}]{gao2011evolution}
\bibinfo{author}{Gao, Q.}, \bibinfo{year}{2011}.
\newblock \bibinfo{title}{Evolution of eddies and packets in turbulent boundary
  layers}.
\newblock Ph.D. thesis. University of Minnesota.
\bibitem[{Gondret et~al.(2002)Gondret, Lance and Petit}]{gondret2002bouncing}
\bibinfo{author}{Gondret, P.}, \bibinfo{author}{Lance, M.},
  \bibinfo{author}{Petit, L.}, \bibinfo{year}{2002}.
\newblock \bibinfo{title}{Bouncing motion of spherical particles in fluids}.
\newblock \bibinfo{journal}{Physics of Fluids} \bibinfo{volume}{14},
  \bibinfo{pages}{643--652}.
\bibitem[{Hall(1988)}]{hall_1988}
\bibinfo{author}{Hall, D.}, \bibinfo{year}{1988}.
\newblock \bibinfo{title}{Measurements of the mean force on a particle near a
  boundary in turbulent flow}.
\newblock \bibinfo{journal}{Journal of Fluid Mechanics} \bibinfo{volume}{187},
  \bibinfo{pages}{451–466}.
\bibitem[{van Hout(2013)}]{van2013spatially}
\bibinfo{author}{van Hout, R.}, \bibinfo{year}{2013}.
\newblock \bibinfo{title}{Spatially and temporally resolved measurements of
  bead resuspension and saltation in a turbulent water channel flow}.
\newblock \bibinfo{journal}{Journal of Fluid Mechanics} \bibinfo{volume}{715},
  \bibinfo{pages}{389--423}.
\bibitem[{van Hout et~al.(2018)van Hout, Eisma, Elsinga and
  Westerweel}]{van2018experimental}
\bibinfo{author}{van Hout, R.}, \bibinfo{author}{Eisma, J.},
  \bibinfo{author}{Elsinga, G.E.}, \bibinfo{author}{Westerweel, J.},
  \bibinfo{year}{2018}.
\newblock \bibinfo{title}{Experimental study of the flow in the wake of a
  stationary sphere immersed in a turbulent boundary layer}.
\newblock \bibinfo{journal}{Physical Review Fluids} \bibinfo{volume}{3},
  \bibinfo{pages}{024601}.
\bibitem[{Jim{\'e}nez et~al.(2010)Jim{\'e}nez, Hoyas, Simens and
  Mizuno}]{jimenez2010turbulent}
\bibinfo{author}{Jim{\'e}nez, J.}, \bibinfo{author}{Hoyas, S.},
  \bibinfo{author}{Simens, M.P.}, \bibinfo{author}{Mizuno, Y.},
  \bibinfo{year}{2010}.
\newblock \bibinfo{title}{Turbulent boundary layers and channels at moderate
  {R}eynolds numbers}.
\newblock \bibinfo{journal}{Journal of Fluid Mechanics} \bibinfo{volume}{657},
  \bibinfo{pages}{335--360}.
\bibitem[{Joseph et~al.(2001)Joseph, Zenit, Hunt and
  Rosenwinkel}]{joseph2001particle}
\bibinfo{author}{Joseph, G.G.}, \bibinfo{author}{Zenit, R.},
  \bibinfo{author}{Hunt, M.L.}, \bibinfo{author}{Rosenwinkel, A.M.},
  \bibinfo{year}{2001}.
\newblock \bibinfo{title}{Particle-wall collisions in a viscous fluid}.
\newblock \bibinfo{journal}{Journal of Fluid Mechanics} \bibinfo{volume}{433},
  \bibinfo{pages}{329--346}.
\bibitem[{Kabsch(1976)}]{kabsch1976solution}
\bibinfo{author}{Kabsch, W.}, \bibinfo{year}{1976}.
\newblock \bibinfo{title}{A solution for the best rotation to relate two sets
  of vectors}.
\newblock \bibinfo{journal}{Acta Crystallographica Section A: Crystal Physics,
  Diffraction, Theoretical and General Crystallography} \bibinfo{volume}{32},
  \bibinfo{pages}{922--923}.
\bibitem[{Kabsch(1978)}]{kabsch1978discussion}
\bibinfo{author}{Kabsch, W.}, \bibinfo{year}{1978}.
\newblock \bibinfo{title}{A discussion of the solution for the best rotation to
  relate two sets of vectors}.
\newblock \bibinfo{journal}{Acta Crystallographica Section A: Crystal Physics,
  Diffraction, Theoretical and General Crystallography} \bibinfo{volume}{34},
  \bibinfo{pages}{827--828}.
\bibitem[{Kaftori et~al.(1995a)Kaftori, Hetsroni and
  Banerjee}]{kaftori1995particle}
\bibinfo{author}{Kaftori, D.}, \bibinfo{author}{Hetsroni, G.},
  \bibinfo{author}{Banerjee, S.}, \bibinfo{year}{1995}a.
\newblock \bibinfo{title}{Particle behavior in the turbulent boundary layer.
  {I}. {M}otion, deposition, and entrainment}.
\newblock \bibinfo{journal}{Physics of Fluids} \bibinfo{volume}{7},
  \bibinfo{pages}{1095--1106}.
\bibitem[{Kaftori et~al.(1995b)Kaftori, Hetsroni and
  Banerjee}]{kaftori1995particleb}
\bibinfo{author}{Kaftori, D.}, \bibinfo{author}{Hetsroni, G.},
  \bibinfo{author}{Banerjee, S.}, \bibinfo{year}{1995}b.
\newblock \bibinfo{title}{Particle behavior in the turbulent boundary layer.
  {II}. {V}elocity and distribution profiles}.
\newblock \bibinfo{journal}{Physics of Fluids} \bibinfo{volume}{7},
  \bibinfo{pages}{1107--1121}.
\bibitem[{Kim et~al.(2014)Kim, Choi, Park and Yoo}]{kim2014inverse}
\bibinfo{author}{Kim, J.}, \bibinfo{author}{Choi, H.}, \bibinfo{author}{Park,
  H.}, \bibinfo{author}{Yoo, J.Y.}, \bibinfo{year}{2014}.
\newblock \bibinfo{title}{Inverse {M}agnus effect on a rotating sphere: when
  and why}.
\newblock \bibinfo{journal}{Journal of Fluid Mechanics} \bibinfo{volume}{754},
  \bibinfo{pages}{R2}.
\bibitem[{Klein et~al.(2013)Klein, Gibert, B{\'e}rut and
  Bodenschatz}]{klein2013simultaneous}
\bibinfo{author}{Klein, S.}, \bibinfo{author}{Gibert, M.},
  \bibinfo{author}{B{\'e}rut, A.}, \bibinfo{author}{Bodenschatz, E.},
  \bibinfo{year}{2013}.
\newblock \bibinfo{title}{Simultaneous 3{D} measurement of the translation and
  rotation of finite-size particles and the flow field in a fully developed
  turbulent water flow}.
\newblock \bibinfo{journal}{Measurement Science and Technology}
  \bibinfo{volume}{24}, \bibinfo{pages}{024006}.
\bibitem[{Law and Thompson(2014)}]{law2014microplastics}
\bibinfo{author}{Law, K.L.}, \bibinfo{author}{Thompson, R.C.},
  \bibinfo{year}{2014}.
\newblock \bibinfo{title}{Microplastics in the seas}.
\newblock \bibinfo{journal}{Science} \bibinfo{volume}{345},
  \bibinfo{pages}{144--145}.
\bibitem[{Loth(2008)}]{loth2008lift}
\bibinfo{author}{Loth, E.}, \bibinfo{year}{2008}.
\newblock \bibinfo{title}{Lift of a solid spherical particle subject to
  vorticity and/or spin}.
\newblock \bibinfo{journal}{AIAA Journal} \bibinfo{volume}{46},
  \bibinfo{pages}{801--809}.
\bibitem[{Magnus(1853)}]{magnus1853ueber}
\bibinfo{author}{Magnus, G.}, \bibinfo{year}{1853}.
\newblock \bibinfo{title}{Ueber die abweichung der geschosse, und: Ueber eine
  auffallende erscheinung bei rotirenden k{\"o}rpern}.
\newblock \bibinfo{journal}{Annalen der Physik} \bibinfo{volume}{164},
  \bibinfo{pages}{1--29}.
\bibitem[{Mathai et~al.(2016)Mathai, Neut, van~der Poel and
  Sun}]{mathai2016translational}
\bibinfo{author}{Mathai, V.}, \bibinfo{author}{Neut, M.W.M.},
  \bibinfo{author}{van~der Poel, E.P.}, \bibinfo{author}{Sun, C.},
  \bibinfo{year}{2016}.
\newblock \bibinfo{title}{Translational and rotational dynamics of a large
  buoyant sphere in turbulence}.
\newblock \bibinfo{journal}{Experiments in Fluids} \bibinfo{volume}{57},
  \bibinfo{pages}{51}.
\bibitem[{Mollinger and Nieuwstadt(1996)}]{mollinger1996measurement}
\bibinfo{author}{Mollinger, A.M.}, \bibinfo{author}{Nieuwstadt, F.T.M.},
  \bibinfo{year}{1996}.
\newblock \bibinfo{title}{Measurement of the lift force on a particle fixed to
  the wall in the viscous sublayer of a fully developed turbulent boundary
  layer}.
\newblock \bibinfo{journal}{Journal of Fluid Mechanics} \bibinfo{volume}{316},
  \bibinfo{pages}{285--306}.
\bibitem[{Monty et~al.(2009)Monty, Hutchins, Ng, Marusic and
  Chong}]{monty2009comparison}
\bibinfo{author}{Monty, J.P.}, \bibinfo{author}{Hutchins, N.},
  \bibinfo{author}{Ng, H.C.H.}, \bibinfo{author}{Marusic, I.},
  \bibinfo{author}{Chong, M.S.}, \bibinfo{year}{2009}.
\newblock \bibinfo{title}{A comparison of turbulent pipe, channel and boundary
  layer flows}.
\newblock \bibinfo{journal}{Journal of Fluid Mechanics} \bibinfo{volume}{632},
  \bibinfo{pages}{431--442}.
\bibitem[{Ni{\~n}o and Garc{\'\i}a(1994)}]{nino1994gravel2}
\bibinfo{author}{Ni{\~n}o, Y.}, \bibinfo{author}{Garc{\'\i}a, M.H.},
  \bibinfo{year}{1994}.
\newblock \bibinfo{title}{Gravel saltation: 2. {M}odeling}.
\newblock \bibinfo{journal}{Water Resources Research} \bibinfo{volume}{30},
  \bibinfo{pages}{1915--1924}.
\bibitem[{Ni{\~n}o and Garc{\'\i}a(1996)}]{ninto1996experiments}
\bibinfo{author}{Ni{\~n}o, Y.}, \bibinfo{author}{Garc{\'\i}a, M.H.},
  \bibinfo{year}{1996}.
\newblock \bibinfo{title}{Experiments on particle-turbulence interactions in
  the near-wall region of an open channel flow: implications for sediment
  transport}.
\newblock \bibinfo{journal}{Journal of Fluid Mechanics} \bibinfo{volume}{326},
  \bibinfo{pages}{285--319}.
\bibitem[{Ni{\~n}o et~al.(1994)Ni{\~n}o, Garc{\'\i}a and
  Ayala}]{nino1994gravel}
\bibinfo{author}{Ni{\~n}o, Y.}, \bibinfo{author}{Garc{\'\i}a, M.H.},
  \bibinfo{author}{Ayala, L.}, \bibinfo{year}{1994}.
\newblock \bibinfo{title}{Gravel saltation: 1. {E}xperiments}.
\newblock \bibinfo{journal}{Water Resources Research} \bibinfo{volume}{30},
  \bibinfo{pages}{1907--1914}.
\bibitem[{Olsen et~al.(1982)Olsen, Cutshall and Larsen}]{olsen1982pollutant}
\bibinfo{author}{Olsen, C.R.}, \bibinfo{author}{Cutshall, N.H.},
  \bibinfo{author}{Larsen, I.L.}, \bibinfo{year}{1982}.
\newblock \bibinfo{title}{Pollutant-particle associations and dynamics in
  coastal marine environments: a review}.
\newblock \bibinfo{journal}{Marine Chemistry} \bibinfo{volume}{11},
  \bibinfo{pages}{501--533}.
\bibitem[{Pan and Banerjee(1997)}]{pan1997numerical}
\bibinfo{author}{Pan, Y.}, \bibinfo{author}{Banerjee, S.},
  \bibinfo{year}{1997}.
\newblock \bibinfo{title}{Numerical investigation of the effects of large
  particles on wall-turbulence}.
\newblock \bibinfo{journal}{Physics of Fluids} \bibinfo{volume}{9},
  \bibinfo{pages}{3786--3807}.
\bibitem[{Pedinotti et~al.(1992)Pedinotti, Mariotti and
  Banerjee}]{pedinotti1992direct}
\bibinfo{author}{Pedinotti, S.}, \bibinfo{author}{Mariotti, G.},
  \bibinfo{author}{Banerjee, S.}, \bibinfo{year}{1992}.
\newblock \bibinfo{title}{Direct numerical simulation of particle behaviour in
  the wall region of turbulent flows in horizontal channels}.
\newblock \bibinfo{journal}{International Journal of Multiphase Flow}
  \bibinfo{volume}{18}, \bibinfo{pages}{927--941}.
\bibitem[{Peng et~al.(2019)Peng, Ayala and Wang}]{peng2019direct}
\bibinfo{author}{Peng, C.}, \bibinfo{author}{Ayala, O.M.},
  \bibinfo{author}{Wang, L.P.}, \bibinfo{year}{2019}.
\newblock \bibinfo{title}{A direct numerical investigation of two-way
  interactions in a particle-laden turbulent channel flow}.
\newblock \bibinfo{journal}{Journal of Fluid Mechanics} \bibinfo{volume}{875},
  \bibinfo{pages}{1096--1144}.
\bibitem[{Pope(2000)}]{pope_2000}
\bibinfo{author}{Pope, S.B.}, \bibinfo{year}{2000}.
\newblock \bibinfo{title}{Turbulent Flows}.
\newblock \bibinfo{publisher}{Cambridge University Press}.
\bibitem[{Rashidi et~al.(1990)Rashidi, Hetsroni and
  Banerjee}]{rashidi1990particle}
\bibinfo{author}{Rashidi, M.}, \bibinfo{author}{Hetsroni, G.},
  \bibinfo{author}{Banerjee, S.}, \bibinfo{year}{1990}.
\newblock \bibinfo{title}{Particle-turbulence interaction in a boundary layer}.
\newblock \bibinfo{journal}{International Journal of Multiphase Flow}
  \bibinfo{volume}{16}, \bibinfo{pages}{935--949}.
\bibitem[{Saffman(1965)}]{saffman1965lift}
\bibinfo{author}{Saffman, P.G.}, \bibinfo{year}{1965}.
\newblock \bibinfo{title}{The lift on a small sphere in a slow shear flow}.
\newblock \bibinfo{journal}{Journal of Fluid Mechanics} \bibinfo{volume}{22},
  \bibinfo{pages}{385--400}.
\bibitem[{Schneiders and Sciacchitano(2017)}]{schneiders2017track}
\bibinfo{author}{Schneiders, J.F.G.}, \bibinfo{author}{Sciacchitano, A.},
  \bibinfo{year}{2017}.
\newblock \bibinfo{title}{Track benchmarking method for uncertainty
  quantification of particle tracking velocimetry interpolations}.
\newblock \bibinfo{journal}{Measurement Science and Technology}
  \bibinfo{volume}{28}, \bibinfo{pages}{065302}.
\bibitem[{Shi and Rzehak(2019)}]{shi2019lift}
\bibinfo{author}{Shi, P.}, \bibinfo{author}{Rzehak, R.}, \bibinfo{year}{2019}.
\newblock \bibinfo{title}{Lift forces on solid spherical particles in unbounded
  flows}.
\newblock \bibinfo{journal}{Chemical Engineering Science}
  \bibinfo{volume}{208}, \bibinfo{pages}{115145}.
\bibitem[{Soldati and Marchioli(2009)}]{soldati2009physics}
\bibinfo{author}{Soldati, A.}, \bibinfo{author}{Marchioli, C.},
  \bibinfo{year}{2009}.
\newblock \bibinfo{title}{Physics and modelling of turbulent particle
  deposition and entrainment: Review of a systematic study}.
\newblock \bibinfo{journal}{International Journal of Multiphase Flow}
  \bibinfo{volume}{35}, \bibinfo{pages}{827--839}.
\bibitem[{Sumer and Deigaard(1981)}]{sumer1981particle}
\bibinfo{author}{Sumer, B.M.}, \bibinfo{author}{Deigaard, R.},
  \bibinfo{year}{1981}.
\newblock \bibinfo{title}{Particle motions near the bottom in turbulent flow in
  an open channel. {P}art 2}.
\newblock \bibinfo{journal}{Journal of Fluid Mechanics} \bibinfo{volume}{109},
  \bibinfo{pages}{311--337}.
\bibitem[{Sumer and O{\u{g}}uz(1978)}]{sumer1978particle}
\bibinfo{author}{Sumer, B.M.}, \bibinfo{author}{O{\u{g}}uz, B.},
  \bibinfo{year}{1978}.
\newblock \bibinfo{title}{Particle motions near the bottom in turbulent flow in
  an open channel}.
\newblock \bibinfo{journal}{Journal of Fluid Mechanics} \bibinfo{volume}{86},
  \bibinfo{pages}{109--127}.
\bibitem[{Sutherland(1967)}]{sutherland1967proposed}
\bibinfo{author}{Sutherland, A.J.}, \bibinfo{year}{1967}.
\newblock \bibinfo{title}{Proposed mechanism for sediment entrainment by
  turbulent flows}.
\newblock \bibinfo{journal}{Journal of Geophysical Research}
  \bibinfo{volume}{72}, \bibinfo{pages}{6183--6194}.
\bibitem[{Tee et~al.(2019)Tee, Barros and Longmire}]{tee2019threedimensional}
\bibinfo{author}{Tee, Y.H.}, \bibinfo{author}{Barros, D.},
  \bibinfo{author}{Longmire, E.K.}, \bibinfo{year}{2019}.
\newblock \bibinfo{title}{Three-dimensional tracking of finite-size spheres in
  a turbulent boundary layer}, in: \bibinfo{booktitle}{Proceedings of the
  13\textsuperscript{th} International Symposium on Particle Image Velocimetry.
  Munich, Germany}, pp. \bibinfo{pages}{740--749}.
\bibitem[{Wallace et~al.(1972)Wallace, Eckelmann and Brodkey}]{wallace1972wall}
\bibinfo{author}{Wallace, J.M.}, \bibinfo{author}{Eckelmann, H.},
  \bibinfo{author}{Brodkey, R.S.}, \bibinfo{year}{1972}.
\newblock \bibinfo{title}{The wall region in turbulent shear flow}.
\newblock \bibinfo{journal}{Journal of Fluid Mechanics} \bibinfo{volume}{54},
  \bibinfo{pages}{39--48}.
\bibitem[{Westerweel and Scarano(2005)}]{westerweel2005universal}
\bibinfo{author}{Westerweel, J.}, \bibinfo{author}{Scarano, F.},
  \bibinfo{year}{2005}.
\newblock \bibinfo{title}{Universal outlier detection for {PIV} data}.
\newblock \bibinfo{journal}{Experiments in Fluids} \bibinfo{volume}{39},
  \bibinfo{pages}{1096--1100}.
\bibitem[{White and Schulz(1977)}]{white1977magnus}
\bibinfo{author}{White, B.R.}, \bibinfo{author}{Schulz, J.C.},
  \bibinfo{year}{1977}.
\newblock \bibinfo{title}{Magnus effect in saltation}.
\newblock \bibinfo{journal}{Journal of Fluid Mechanics} \bibinfo{volume}{81},
  \bibinfo{pages}{497--512}.
\bibitem[{Wieneke(2008)}]{wieneke2008volume}
\bibinfo{author}{Wieneke, B.}, \bibinfo{year}{2008}.
\newblock \bibinfo{title}{Volume self-calibration for 3{D} particle image
  velocimetry}.
\newblock \bibinfo{journal}{Experiments in Fluids} \bibinfo{volume}{45},
  \bibinfo{pages}{549--556}.
\bibitem[{Yousefi et~al.(2020)Yousefi, Costa and Brandt}]{yousefi2020single}
\bibinfo{author}{Yousefi, A.}, \bibinfo{author}{Costa, P.},
  \bibinfo{author}{Brandt, L.}, \bibinfo{year}{2020}.
\newblock \bibinfo{title}{Single sediment dynamics in turbulent flow over a
  porous bed-insights from interface-resolved simulations}.
\newblock \bibinfo{journal}{Journal of Fluid Mechanics} \bibinfo{volume}{893},
  \bibinfo{pages}{A24, 1--28}.
\bibitem[{Zeng et~al.(2008)Zeng, Balachandar, Fischer and
  Najjar}]{zeng2008interactions}
\bibinfo{author}{Zeng, L.}, \bibinfo{author}{Balachandar, S.},
  \bibinfo{author}{Fischer, P.}, \bibinfo{author}{Najjar, F.},
  \bibinfo{year}{2008}.
\newblock \bibinfo{title}{Interactions of a stationary finite-sized particle
  with wall turbulence}.
\newblock \bibinfo{journal}{Journal of Fluid Mechanics} \bibinfo{volume}{594},
  \bibinfo{pages}{271--305}.
\bibitem[{Zhao and Andersson(2011)}]{zhao2011particle}
\bibinfo{author}{Zhao, L.}, \bibinfo{author}{Andersson, H.I.},
  \bibinfo{year}{2011}.
\newblock \bibinfo{title}{On particle spin in two-way coupled turbulent channel
  flow simulations}.
\newblock \bibinfo{journal}{Physics of Fluids} \bibinfo{volume}{23},
  \bibinfo{pages}{093302}.
\bibitem[{Zimmermann et~al.(2011)Zimmermann, Gasteuil, Bourgoin, Volk, Pumir
  and Pinton}]{zimmermann2011tracking}
\bibinfo{author}{Zimmermann, R.}, \bibinfo{author}{Gasteuil, Y.},
  \bibinfo{author}{Bourgoin, M.}, \bibinfo{author}{Volk, R.},
  \bibinfo{author}{Pumir, A.}, \bibinfo{author}{Pinton, J.F.},
  \bibinfo{year}{2011}.
\newblock \bibinfo{title}{Tracking the dynamics of translation and absolute
  orientation of a sphere in a turbulent flow}.
\newblock \bibinfo{journal}{Review of Scientific Instruments}
  \bibinfo{volume}{82}, \bibinfo{pages}{033906}.

\end{thebibliography}

\end{document}